\documentclass[hidelinks]{article}
\usepackage[total={7in,9in}]{geometry}
\usepackage{hyperref}
\usepackage{rotating}

\usepackage{makecell,multirow}
\usepackage{float}
\usepackage{algorithm,amsmath,amsthm}
\usepackage{algpseudocode}
\usepackage{xcolor}
\usepackage[compact]{titlesec}

\usepackage[round]{natbib}
\usepackage{caption}
\usepackage{graphicx}
\usepackage{bm}
\usepackage{authblk}

\providecommand{\keywords}[1]{\textbf{\textit{Keywords:}} #1}

\DeclareMathOperator*{\argmax}{arg\,max}
\DeclareMathOperator*{\pr}{Pr}
\DeclareMathOperator{\Var}{Var}

\DeclareMathOperator{\ber}{Bernoulli}
\DeclareMathOperator{\logit}{logit}

\newcommand{\groupR}{G}
\newcommand{\group}{g}

\newcolumntype{s}{>{\hsize=0.6\hsize}X}
\newcolumntype{q}{>{\hsize=1.4\hsize}X}
\algnewcommand\INPUT{\item[\textbf{Input:}]}%
\algnewcommand\OUTPUT{\item[\textbf{Output:}]}%
\algnewcommand\function{\item[\textbf{function}]}%
\newtheorem{property}{Property}
\theoremstyle{definition}
\newtheorem{definition}{Definition}[section]

\begin{document}

\title{Incorporating Participants' Welfare into Sequential Multiple Assignment Randomized Trials}

\small
\author[1]{Xinru Wang}
\author[2,3]{Nina Deliu}
\author[4]{Yusuke Narita}
\author[1,5,6,7$\ast$]{Bibhas Chakraborty}

\footnotesize{
\affil[1]{Centre for Quantitative Medicine, Duke-NUS Medical School, Singapore} 
\affil[2]{MRC - Biostatistics Unit, University of Cambridge, Cambridge, UK} 
\affil[3]{Department of Methods and Models for Economics, Territory and Finance, Sapienza University of Rome, Rome, Italy} 
\affil[4]{Department of Economics and Cowles Foundation, Yale University, New Haven, Connecticut, USA} 
\affil[5]{Program in Health Services and Systems Research, Duke-NUS Medical School, Singapore} 
\affil[6]{Department of Statistics and Data Science, National University of Singapore, Singapore} 
\affil[7]{Department of Biostatistics and Bioinformatics, Duke University, Durham, North Carolina, USA} 
}

\date{}

\maketitle

\begin{abstract}
Dynamic treatment regimes (DTRs) are sequences of decision rules that recommend treatments based on patients' time-varying clinical conditions. The sequential multiple assignment randomized trial (SMART) is an experimental design that can provide high-quality evidence for constructing optimal DTRs. In a conventional SMART, participants are randomized to available treatments at multiple stages with balanced randomization probabilities. Despite its relative simplicity of implementation and desirable performance in comparing embedded DTRs, the conventional SMART faces inevitable ethical issues including assigning many participants to the empirically inferior treatment or the treatment they dislike, which might slow down the recruitment procedure and lead to higher attrition rates, ultimately leading to poor internal and external validities of the trial results. In this context, we propose a SMART under the Experiment-as-Market framework (SMART-EXAM), a novel SMART design that holds the potential to improve participants' welfare by incorporating their preferences and predicted treatment effects into the randomization procedure. We describe the steps of conducting a SMART-EXAM and evaluate its performance compared to the conventional SMART. The results indicate that the SMART-EXAM can improve the welfare of the participants enrolled in the trial, while also achieving a desirable ability to construct an optimal DTR when the experimental parameters are suitably specified. We finally illustrate the practical potential of the SMART-EXAM design using data from a SMART for children with attention-deficit/hyperactivity disorder (ADHD).
\end{abstract}

\section{Introduction}
\label{sec:section1}

\begin{figure}[htb]
\centering
\includegraphics[width=0.8\linewidth]{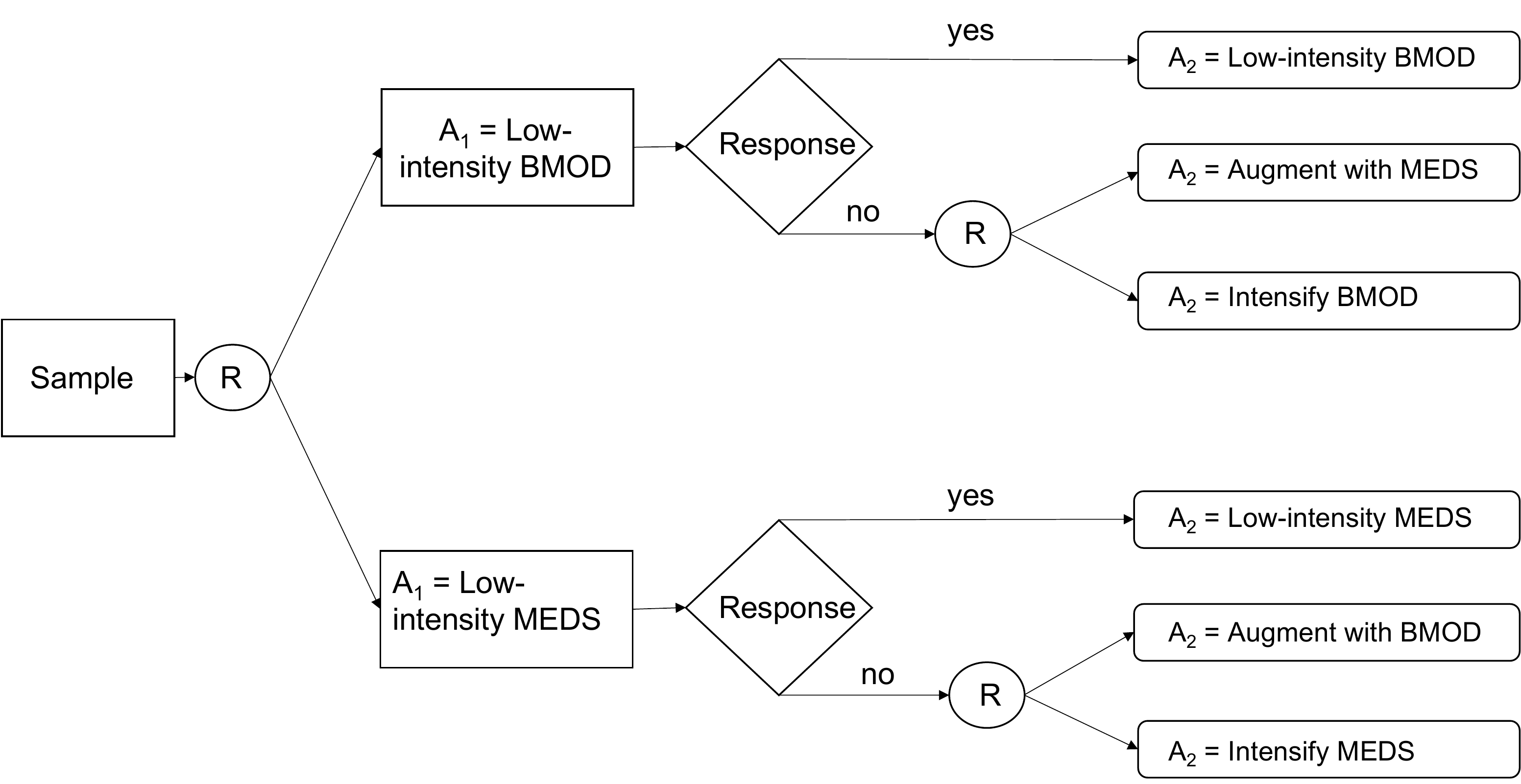}
\caption{An example of a two-stage SMART design for children with ADHD. R denotes randomization. BMOD and MEDS represent behavioral modification and oral methamphetamine respectively. }
\label{smart_2}
\end{figure}

The management of chronic and relapsing diseases requires adjusting treatments at different time points based on previous treatment histories and disease status to improve the final outcomes of interest. Dynamic treatment regimes (DTRs) are sequences of decision rules that can formalize the patient-centered disease management model designed to guide clinicians in prescribing the right treatments to the right patients. The advantages of DTRs over fixed treatments have been recognized by researchers in various health domains, from the perspectives of maximizing treatment effectiveness, minimizing side effects, and improving cost-effectiveness. The sequential multiple assignment randomized trial (SMART), an experimental design with multiple randomization stages, is considered the gold-standard design for constructing optimal DTRs \citep{murphy2005experimental, wang2023sequential}. Figure~\ref{smart_2} illustrates a SMART for children with attention-deficit/hyperactivity disorder (ADHD) \citep{pelham2016treatment}. At stage 1, children with ADHD were randomized to either low-intensity behavioral modification (BMOD) or low-dose oral methamphetamine (MEDS). After eight weeks of treatment, the response status of each child was assessed using the Impairment Rating Scale and an individualized list of target behaviors on a monthly basis. If the children showed insufficient response to the treatment, they were re-randomized to either intensify the initial treatment or augment with another treatment. Responders continued with the initial treatment and got re-randomized if the disease conditions deteriorated at any time after eight weeks. In this SMART, there were four embedded DTRs, one of which is: ``begin with low-intensity BMOD and add MEDS at stage 2 for non-responders.''

Numerous methods have been established to analyze data from SMARTs, with the aim of estimating stage-specific treatment effects, comparing the embedded DTRs, or constructing more tailored DTRs \citep{oetting2007statistical, nahum2012q}. From the design perspective, the conventional SMART (henceforth referred to as the SMART where there is no confusion), i.e., SMART with balanced randomization probabilities across treatments at each stage, can maximize the ability to compare the embedded DTRs \citep{murphy2005experimental}. However, with various data sources available, e.g., electronic health records, cohort studies, patient databases, and previous trials, for estimating treatment effects, it has become debatable whether it is ethical to evenly randomize participants without considering the potential individualized treatment effects. Furthermore, in recent years, a broader trend towards patient-centered care has emerged, with a growing emphasis on patient preferences and \textit{shared decision making} \citep{fukami2023shared}. Under this framework, healthcare providers and patients exchange information about patients' disease-related characteristics and the benefits and side effects of each treatment, and finally achieve a consensus about which treatment(s) should be prescribed. This paradigm shift is grounded in the principle of patient autonomy, with the aim of promoting patients' participation in healthcare decision-making. In view of this paradigm shift, it may be unreasonable to evenly randomize participants ignoring their preferences in SMARTs altogether. Particularly noteworthy is the fact that the treatment components in a SMART are often available in routine care which may make participants hesitate to join the trial due to concerns about lower probabilities of receiving their preferred treatment by entering into the trial \citep{edwards2000can}.  Ignoring the above issues may exert a negative impact on the recruitment procedure and demotivate participants from sticking to their assigned treatments, which may ultimately reduce the power and internal and external validities of the trial results \citep{gaines2011experimental}. 

The above ethical issues are also present in randomized controlled trials (RCTs). The response-adaptive randomization (RAR) \citep{eisele1994doubly, rosenberger2004maximizing, david2023response} was proposed to allow for potential adjustments of randomization probabilities, in accordance with accumulating information about the treatment performance. In addition, a lot of work has been done to either incorporate participants' preferences during randomization or estimate average choice-specific treatment effects (ACTE), i.e., the conditional treatment effect among those who would choose a particular treatment option. Specifically, two-stage randomized designs \citep{rucker1989two} first randomize participants into either the randomization group or the choice group. Participants in the randomization group will be randomized again to the available treatments, while participants in the choice group will receive their preferred treatment. By randomizing twice, such designs enable researchers to estimate ACTE \citep{knox2019design} and increase the probability of allocating participants to their preferred treatment as compared to conventional RCTs. In addition, fully randomized preference designs \citep{torgerson1996patient} first elicit participants' preferences before randomization, and then use the elicited preferences to estimate ACTE. One of their limitations is that there may be discrepancies between participants' stated preferences and actual choices, and it may be unethical to continue with balanced randomization probabilities after asking about their preferences. In partially randomized preference designs, participants with strong preferences are given their preferred treatment, while those with mild preferences are randomized to available treatments. While such designs have the potential to recruit more participants with strong preferences who would have refused to participate had they got randomized, the estimated average treatment effects (ATEs) are restricted among those with mild preferences instead of a wider population \citep{walter2017beyond}. 

There have been several SMARTs that incorporated participants' preferences during the randomization procedure \citep{fava2003background, mckay2015effect}. The Sequenced Treatment Alternatives to Relieve Depression (STAR*D) was a multisite, multi-stage, first-generation SMART for patients with major depressive disorder \citep{fava2003background}. In STAR*D, participants were randomized within their preferred treatment categories (switch or augment) during the second and third stages. From the analysis perspective, there have been some works \citep{wang2012evaluation, johnson2021optimizing, xu2022estimating} to construct composite outcomes accounting for both treatment efficacy and other factors, e.g., toxicity and quality of life, to evaluate DTRs. The parameters in the outcome regarding these factors may be chosen based on patients' preferences, e.g., whether patients prefer higher treatment effects, lower side effects, or lower costs. All of these works demonstrate the paramount importance of incorporating preferences in SMARTs.

Furthermore, attempts have been made to incorporate predicted treatment effects in SMARTs. \citet{cheung2015sequential} proposed a SMART with adaptive randomization (SMART-AR) using Q-learning to determine the randomization probabilities in favor of superior treatments based on the complete data trajectories from previous participants in the trial. \citet{wang2021adaptive} introduced a response-adaptive SMART (RA-SMART) that employs a framework akin to the ``play the winner'' rule, i.e., the inferior treatment at stage 1 will have a lower randomization probability at stage 2. \citet{wu2023interim} presented a SMART with interim monitoring, where early termination is permitted if there is sufficient evidence of treatment efficacy. Nevertheless, none of the aforementioned designs simultaneously incorporated participants' preferences and treatment effects into the randomization procedure. 

Recently, \citet{narita2021incorporating} proposed an extension of the RCT --- the Experiment-as-Market (EXAM) design, which allows for incorporating both participants' preferences and treatment effects into the randomization procedure, while maintaining robust inferences comparable to RCTs. Inspired by this, the current paper proposes a novel SMART design under the Experiment-as-Market framework (SMART-EXAM), which has the potential to improve participants' welfare in terms of both the participants' preferences and their final outcomes. By taking treatment effects and participants' preferences into account to individualize the randomization probabilities, the SMART-EXAM mimics a shared decision-making process between patients and healthcare providers in a clinical trial setting, which enables participants to actively participate in the randomization procedure. 

The remainder of the article is divided into five sections. \hyperref[sec:section2]{Section 2} illustrates the structure of a generic two-stage SMART used throughout this paper and presents analysis methods to compare embedded DTRs in a SMART. \hyperref[sec:section3]{Section 3}  presents the detailed procedure of conducting a SMART-EXAM and its theoretical properties. In \hyperref[sec:section4]{Section 4}, we conduct a simulation study to evaluate the empirical performance of the SMART-EXAM compared to the SMART. In \hyperref[sec:section5]{Section 5}, we demonstrate the practical potential of the SMART-EXAM using data from a SMART ADHD study. The last section comes with conclusions and discussions about the SMART-EXAM.

\section{Set up and Notation}
\label{sec:section2}
To facilitate the exposition, we focus on a two-stage SMART that is consistent with the SMART ADHD study \citep{pelham2016treatment}. The observed trajectory for the $i$-th participant is denoted by $(\bm{O}_{1i}, A_{1i}, \bm{O}_{2i}, A_{2i}, Y_i)$, which is assumed to be independent and identically distributed. $\bm{O}_{ti}\ (t = 1, 2 )$ denotes the vector of covariates obtained prior to treatment at stage $t$. Among covariates $\bm{O}_{2i}$, $R_i$ is the indicator for intermediate response status, with $R_i=1/0$ for responders and non-responders, respectively. $A_{ti}$ is the treatment indicator at stage $t$, with  $A_{1i} \in \{-1, 1 \}$, $A_{2i} \in \{-1, 1 \}$ for non-responders, and $A_{2i} = 0 $ for responders. $Y_i$ is the final continuous outcome and without loss of generality, we assume that higher values of $Y_i$ are preferred. Define $\bm{H}_{ti} \in \mathcal{H}_t$ as the history data of the $i$-th participant at stage $t$, with $\bm{H}_{1i}=\bm{O}_{1i}$, $\bm{H}_{2i}=(\bm{O}_{1i}, A_{1i}, \bm{O}_{2i})$, where $\mathcal{H}_t$ is the space of possible histories at stage $t$. Under the SMART shown in Figure~\ref{smart_2}, the $j$-th two-stage DTR in the space of DTRs of interest $\bm{\mathcal{D}}$ is denoted as $\bm{d}_j = (d_{j1}, d_{j2}), j=1,\dots,J$, where $d_{j1}, d_{j2} \in \{-1,1\}$, and $J$ is the total number of embedded DTRs. For example, $\bm{d}_1 = (1,1)$ means that, first treat participants with $A_1=1$, if they do not respond, switch to $A_2=1$, otherwise continue with the initial treatment. Under the Neyman-Rubin causal inference framework \citep{rubin1974estimating}, the potential outcome under DTR $\bm{d}_j=(d_{j1}, d_{j2})$ is denoted as
\begin{equation}
\label{dtr_potential}
    Y^{\bm{d}_j} = R^{d_{j1}}Y^{s(d_{j1},0)}+(1-R^{d_{j1}})Y^{s(d_{j1},d_{j2})},
\end{equation}
where $s(d_{j1},0)$ denotes the treatment sequence of receiving $A_1=d_{j1}$, responding, and continuing with the initial treatment, while $s(d_{j1},d_{j2})$ denotes the treatment sequence of receiving $A_1=d_{j1}$, not responding, and switching to $A_2=d_{j2}$. Here we use the prefix $s$ to differentiate the notations for DTR $\bm{d}_j$ and the treatment sequence. $Y^{s(d_{j1},0)}$ and $Y^{s(d_{j1},d_{j2})}$ are the potential outcomes for participants with treatment sequences $s(d_{j1},0)$ and $s(d_{j1},d_{j2})$, respectively. The expected outcome of DTR $\bm{d}_j$ is $\mu_{\bm{d}_j} = \ E(Y^{\bm{d}_j}) = \pi_{d_{j1}} \mu_{s(d_{j1},0)} + (1- \pi_{d_{j1}})  \mu_{s(d_{j1}, d_{j2})}$, where $\pi_{d_{j1}}$ is the response rate for treatment $d_{j1}$ at stage 1; $\mu_{s(d_{j1},0)}$ and $\mu_{s(d_{j1}, d_{j2})}$ are the expected outcomes of those with treatment sequences $s(d_{j1},0)$ and $s(d_{j1}, d_{j2})$, respectively. The optimal DTR is defined as the DTR with the highest expected outcome $\mu_{\bm{d}_j}$, i.e., $\bm{d}^{\ast} = \underset{\bm{d}_j \in \bm{\mathcal{D}}}{\argmax}\ \mu_{\bm{d}_j}$.

In this paper, the primary goal is to select an optimal DTR $\bm{d}^{\ast}$ among the embedded DTRs in a SMART. Although there are various methods for estimating DTR means, e.g., G-computation and augmented inverse probability weighting (AIPW), we mainly focus on the inverse probability weighting (IPW) due to both its unbiasedness and ease of implementation in the context of the SMART-EXAM. Under usual causal assumptions in the Neyman-Rubin causal framework, the IPW estimator for the expected outcome of DTR $\bm{d}_j = (d_{j1},d_{j2})$ is defined as
\begin{equation}
\label{ipw_1}
\hat{\mu}_{\bm{d}_j} = \frac{\sum_{i=1}^N W^{\bm{d}_j}_i Y_i }{\sum_{i=1}^N W^{\bm{d}_j}_i} ,
\end{equation}
where $N$ is the total sample size, $W^{\bm{d}_j}_i= \frac{I(A_{1i}=d_{j1})}{p_{1,d_{j1},i}} \left\{R_i+ \frac{(1-R_i)I(A_{2i}=d_{j2})}{p_{2,d_{j2},i}} \right\}$ is the weight of the $i$-th individual for DTR $\bm{d}_j = (d_{j1}, d_{j2})$, $p_{1,d_{j1},i} = \pr(A_{1i}=d_{j1})$, and $p_{2,d_{j2},i} = \pr(A_{2i}= d_{j2}|R_i=0,A_{1i}=d_{j1})$. The variance estimator of the DTR mean estimator \citep{oetting2007statistical} is 
\begin{equation}
\label{ipw_3}
\widehat{\Var}(\hat{\mu}_{\bm{d}_j}) =\frac{\sum_{i=1}^N (W^{\bm{d}_j}_i(Y_i-\hat{\mu}_{\bm{d}_j}))^2 }{N^2}.
\end{equation}
The estimated optimal DTR $\hat{\bm{d}}^{\ast}$ is thus defined as $\hat{\bm{d}}^{\ast} = \underset{\bm{d}_j}{\argmax}\ \hat{\mu}_{\bm{d}_j}$. 

\section{SMART with the Experiment-as-Market Framework (SMART-EXAM)}
\label{sec:section3}
As an extension of RCTs, the EXAM aims to strike a balance between making accurate inferences for future patients and improving the welfare of enrolled participants through an imaginary centralized market \citep{narita2021incorporating}. One can view the EXAM as a shared decision-making randomization procedure where healthcare providers rely on existing evidence about treatment effects to determine the most effective treatment, while participants indicate their preferences based on previous treatment experience and information provided by professionals about potential benefits and side effects of each treatment. The EXAM can generate randomization probabilities that take into account the perspectives of both healthcare providers and participants and achieve a so-called consensus about the randomization probabilities. It has been proved that the EXAM can improve participants' welfare while ensuring valid inferences about the ATEs, on the grounds that the EXAM can be seen as a special case of stratified randomized trials based on predicted treatment effects and preferences. 

Building upon the work of \citet{narita2021incorporating}, the current paper proposes a novel SMART design - the SMART-EXAM - to simultaneously incorporate participants' preferences and treatment benefits into the randomization procedure. Note that the embedded DTRs in a SMART are often stepped-up treatment strategies, i.e., for non-responders, stepped-up treatments are provided to ``rescue'' the initial treatment. In this case, it is critical to improve the allocation at stage 2. Furthermore, participants tend to have preferences for stage-2 treatments, as is shown in a pilot study for adolescent depression \citep{gunlicks2016pilot}. Against this backdrop, the current paper focuses on SMART-EXAMs with individualized randomization at stage 2, while maintaining balanced randomization at stage 1.

\subsection{Key Definitions}
\label{sec:section3.1}

To facilitate discussion, we first provide basic definitions for a SMART-EXAM. Among non-responders to the initial treatment $a_1$, we define:

\begin{definition}[\textit{Preferences}]
Let $\Lambda_{i} \in \{0,1 \}$ denote the stage-2 preference indicator of the $i$-th participant, with $\Lambda_{i}=1$ and $\Lambda_i=0$ corresponding to the cases where the participant prefers $A_2=1$ and $A_2=-1$, respectively. 
\end{definition}

\begin{definition}[\textit{Treatment capacity} and \textit{Treatment demand}]
Let $C_{a_2|a_1}$ be the capacity for $a_2$ among non-responders to $a_1$, such that $\sum_{a_2 \in \{-1,1 \}} C_{a_2|a_1} = N^{nr}_{a_1}$, where $N^{nr}_{a_1}$ is the number of non-responders to $a_1$. The treatment demand $D_{a_2|a_1}$ is defined as the total number of non-responders to $a_1$ who prefer $a_2$, i.e., $D_{a_2|a_1}=\sum_{i=1}^{N^{nr}_{a_1} }\{ I(a_2=1)\Lambda_i + I(a_2=-1) (1-\Lambda_i) \}$. Define the excess-demand and the oversupplied stage-2 treatment as $a^{e}_2=\{a_2 \in \{-1,1\}: D_{a_2|a_1} \geq C_{a_2|a_1} \} $ and $a^{o}_2=\{a_2 \in \{-1,1\}: D_{a_2|a_1} < C_{a_2|a_1} \}$, respectively.
\end{definition}

The choice of treatment capacities depends on various factors \citep{dumville2006use}. For example, allocating less capacity to the more expensive treatment can reduce trial costs, while allocating higher capacity to the treatment with a higher drop-out rate and variance can enhance the overall statistical efficiency. In addition, considerations may also involve assigning more capacity to the treatment on which researchers may be interested in gaining more information, e.g., side effects or treatment experience and assigning more capacity to the treatment that most participants prefer.

\begin{definition}[\textit{Individualized treatment effects}]
Under the Neyman-Rubin causal inference framework, the effect of $A_2=a^{e}_2$ on the outcome compared with $A_2=a^{o}_2$ conditional on the history $\bm{H}_{2i}$ for participants $i=1, \dots, N^{nr}_{a_1}$ is denoted as $\zeta_{2,a_2^e,i} = E[Y_i|\bm{H}_{2i},A_{2i}=a^{e}_2] -  E[Y_i|\bm{H}_{2i},A_{2i}=a^{o}_2] $. For the simplicity of notations, we use $\zeta_{i}$ for such treatment effects where there is no confusion. For example, the outcome model can be specified as
\begin{equation}
\begin{aligned}
\label{qfunction_1}
Q_2(\bm{H}_{2i}, A_{2i}; \bm{\gamma}_2, \bm{\alpha}_2) &= E[Y_i | \bm{H}_{2i}, A_{2i}]  = \bm{\gamma}_2^{T} \bm{H}_{20i}+ \bm{\alpha}_2^{T}\bm{H}_{21i} A_{2i},
\end{aligned}
\end{equation}
where $\bm{H}_{20i}$ and $\bm{H}_{21i}$ are potentially different features of $\bm{H}_{2i}$, and $\bm{H}_{21i}$ is the vector of covariates in $\bm{H}_{2i}$ that are believed to interact with the treatment. The individualized treatment effect $\hat{\zeta}_{i}$ is $2\hat{\bm{\alpha}}^{T}_2 \bm{H}_{21i}$ if $a_2^e=1$, or $-2\hat{\bm{\alpha}}^{T}_2 \bm{H}_{21i}$ if $a_2^e=-1$, where $\hat{\bm{\alpha}}^{T}_2$ is the estimated coefficients based on previous data. Note that when estimating treatment effects using observational data, careful attention should be paid to adjust for potential confounders. In addition to Q-learning with linear regression, other advanced statistical methods, e.g., random forest, regression trees, and boosting, etc. \citep{knaus2021machine}, can be utilized to estimate $\zeta_{i}$. Here we use the model in Eq.~(\ref{qfunction_1}) only for the purpose of illustration.
\end{definition}

Treatment effects and preferences are two complementary measures of participants' welfare, which are sometimes found to be correlated through some common predictors, e.g., age and previous treatment experience. For instance, in the SMART ADHD study, \citet{nahum2012q} found that non-responders with lower adherence have higher effects of augmentation over intensification compared to those with higher adherence. It is also reasonable to assume that non-responders with lower adherence are more likely to prefer augmentation than those with higher adherence. As a result, there is a positive association between treatment effects and preferences. It aligns with the shared decision-making procedure with less trade-off between preferences and treatment effects as perceived by healthcare providers.  

\begin{definition}[\textit{Utility}]
Define the utility of the $i$-th participant as $u_i=p_{2, 1, i}\ \Lambda_{i} + (1-p_{2, 1, i})(1-\Lambda_{i})$.

To appreciate the implications of the utility, we introduce the indicator for whether the $i$-th participant receives the preferred treatment or not, denoted by $K_i=I(A_{2i}=1)\Lambda_i+I(A_{2i}=-1)(1-\Lambda_i)$. The probability of receiving the preferred treatment, i.e., the expected value of $K_i$, conditional on participants' indicated preferences, is thus denoted as $u_i=E[K_i|\Lambda_i]=p_{2, 1, i}\ \Lambda_{i} + (1-p_{2, 1, i})(1-\Lambda_{i})$.
\end{definition}

\textit{Utility} holds significant importance for participants when deciding whether or not to participate in the trial. As stated in \hyperref[sec:section1]{Section 1}, participants may refuse to participate due to concerns about a relatively lower probability of receiving their preferred treatment by entering into the trial. However, some ``altruistic'' participants may accept making small sacrifices and participating to contribute to future patients. The higher the probability of receiving their preferred treatment, the less is altruism required and more participants may agree to participate in SMARTs \citep{edwards2000can}. 

\subsection{Procedure of conducting a SMART-EXAM}
\label{sec:section3.2}

\begin{figure}[htb]
\centering
\includegraphics[width=0.8\linewidth]{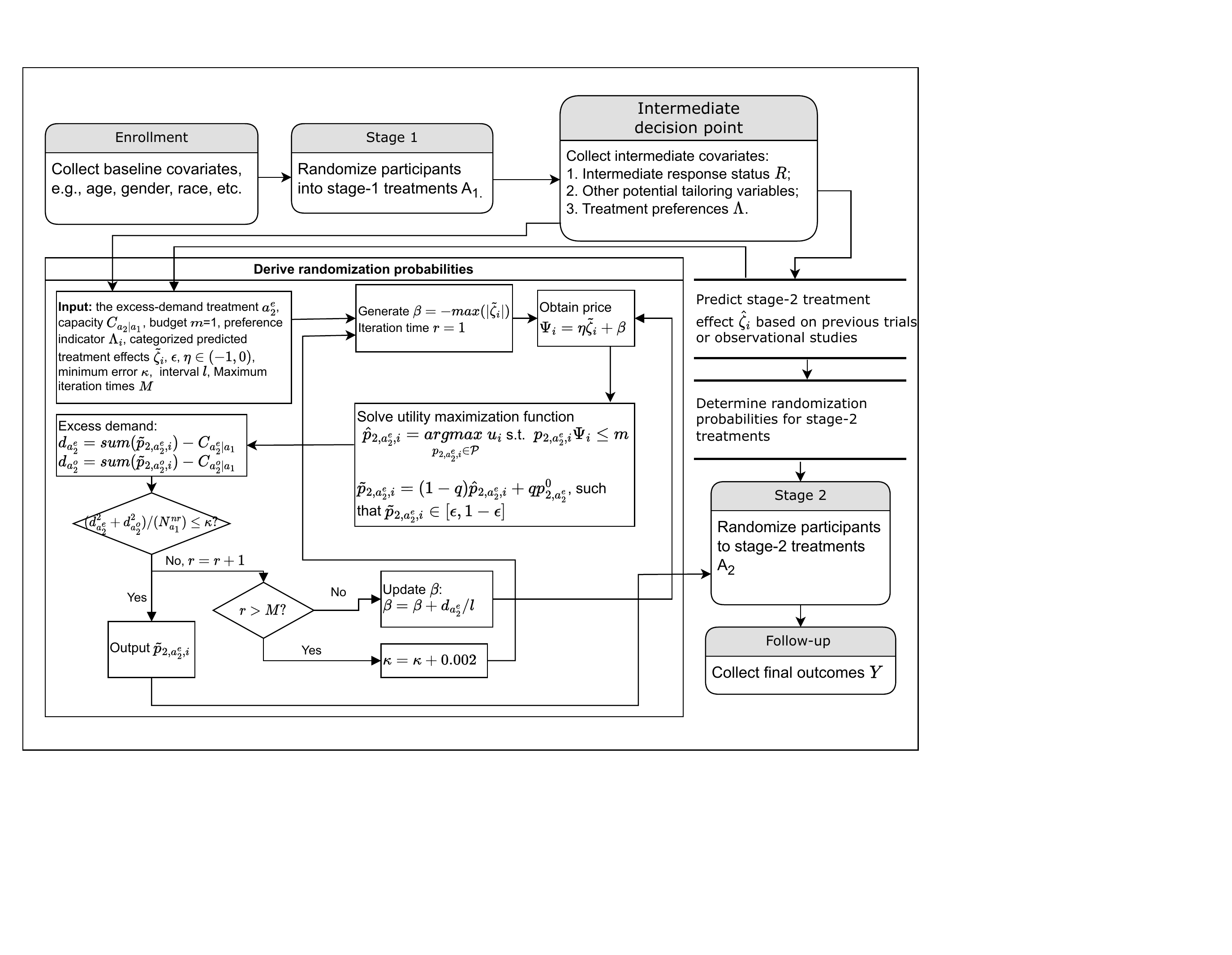}
\caption{The schematic plot for a SMART-EXAM that individualizes the stage-2 randomization probabilities to account for participants' welfare. }
\label{smart-exam}
\end{figure}

Figure~\ref{smart-exam} graphically illustrates a SMART-EXAM that individualizes the stage-2 randomization probabilities, and the corresponding detailed algorithm is available in Web Appendix A. The stage-1 randomization is the same as that in the SMART, while at the intermediate decision point, the preference indicators $\Lambda_i$ are collected and treatment effects $\hat{\zeta}_{i}$ are estimated based on previous trials or observational studies. In the following procedure, the $\hat{\zeta}_{i}$ are normalized so that they have a mean of $0$ and a standard deviation of $1$. To produce less variable randomization probabilities, the normalized treatment effects $\hat{\zeta}_i$ should be discretized by one of the binning strategies, e.g., creating contiguous intervals with equal frequencies and specifying the categorized value of treatment effect $\tilde{\zeta}_i$ as the mean value of the interval. 

Subsequent to estimating treatment effects, each non-responder is given a budget $m =1$. The same budget for each participant is to ensure giving equal importance to every participant’s welfare. Define $\Psi_{i}$ as the treatment price of a unit randomization probability of $a^e_2$ for participant $i$; there exist $\eta \in (-1,0)$ and $\beta \in \mathcal{R}$ such that $\Psi_{i}=\eta \tilde{\zeta}_{i} + \beta$. Note that $\beta$ and $\eta$ are common parameters for the price function; the individualization of price $\Psi_{i}$ is induced by the heterogeneous treatment effect $\tilde{\zeta}_{i}$. Intuitively, $\tilde{\zeta}_{i} < \tilde{\zeta}_{i'}$ corresponds to the cases where the $i'$-th participant will benefit more when $\tilde{\zeta}_{i'} > 0 $ (or lose less when $\tilde{\zeta}_{i'} < 0$) from treatment $a^e_2$ than the $i$-th participant. Thus it is more reasonable to give the $i'$-th participant a higher randomization probability for $a^e_2$ than the $i$-th participant when the treatment resources are limited. This can be accomplished by a negative value of $\eta$, which ensures that, $\Psi_{i} > \Psi_{i'}$ if $\tilde{\zeta}_{i} < \tilde{\zeta}_{i'}$, i.e., the price of a unit randomization probability for $a^e_2$ is lower for those with higher treatment effects, such that they can ``buy'' more randomization probabilities for $a^e_2$ compared to those with lower treatment effects. The values of $\beta$ depend on treatment capacities, which will be described later. 

The randomization probabilities are derived by solving the utility maximization function subject to budget constraints. For participant $i$:
  \begin{equation}
    \label{wtp_2}
  \begin{aligned}
\hat{p}_{2,a_2^e,i} = \underset{ p_{2,a_2^e,i} \in \mathcal{P}}{\argmax} \ u_i \quad s.t. \ p_{2,a_2^e,i}\ \Psi_{i}  \leq m,
      \end{aligned}
  \end{equation}
where $\mathcal{P} \equiv \{p_{2,a_2^e,i} |  p_{2, a_2^e,i}  \in [0,1] \}$, and $p_{2,a_2^e,i}\ \Psi_{i} $ is the expected expense of ``buying'' randomization probabilities for treatment $a_2^e$. The underlying rationale of Eq.~(\ref{wtp_2}) is to maximize the utility, i.e., the probability of receiving the preferred treatment, subject to the budget constraint, with the price determined by the predicted treatment effects. The budget constraint takes predicted treatment effects into the optimization problem and ensures that the limited treatment resources favor those who can benefit more from the treatment. 

To ensure that a non-null subset of participants follow each embedded DTR, i.e., the positivity assumption holds for valid inferences, $\epsilon \in [0,\bar{\epsilon}]$ is introduced to keep the randomization probabilities within the range $[\epsilon, 1-\epsilon]$, where $\bar{\epsilon}=\min_{a_2} p^{0}_{2,a_2}$, and $p^{0}_{2,a_2}=\frac{C_{a_2|a_1}}{N^{nr}_{a_1}}$ is the non-individualized randomization probability for $a_2$ with treatment capacity $C_{a_2|a_1}$. The final stage-2 randomization probability is $\tilde{p}_{2,a_2,i} = (1-q) \hat{p}_{2,a_2,i} + q p^{0}_{2,a_2}$, where $q \equiv \underset{i,a_2}{\inf} \{q^{'} \in [0,1] | (1-q^{'} ) \hat{p}_{2,a_2,i} + q^{'} p^{0}_{2,a_2} \in [\epsilon, 1-\epsilon]\}$. Alternatively, other methods such as trimming the extreme probabilities to a predetermined value can also avoid violating the positivity assumption. The values of $\beta$ in the price function are adjusted through an iterative procedure shown in Figure~\ref{smart-exam}, to ensure that the randomization probabilities derived from the above procedure meet the capacity constraints: $\sum_{i=1}^{N^{nr}_{a_1}}  \tilde{p}_{2, a_2,i} = C_{a_2|a_1}$ for each $a_2 \in \{-1,1\}$, i.e., the expected number of non-responders allocated to $a_2$ under the current randomization mechanism equals to the capacity of $a_2$; thus, the design satisfies a system-level capacity constraint. Note that in Figure~\ref{smart-exam}, interval $l$ can be interpreted as the step size for updating the value of $\beta$. $(d_{a_2^e}^2 +d_{a_2^o}^2)
  / (N^{nr}_{a_1} )$ is compared to $\kappa$ to assess whether the sum of the square differences between the expected number of participants and the capacity for each treatment, i.e., $d^2_{a_2^e} + d^2_{a_2^o}$, is within a desired range $[0,\kappa N^{nr}_{a_1}]$ or not. 

By the above procedure, participants with the same $\Lambda_i$ and $\tilde{\zeta}_i$ will be given the same randomization probabilities as they have the same utility and budget constraint. In other words, non-responders can be divided into $B$ groups indexed by $\groupR_{i} \in \{ 1, \dots, B \}$ based on $\Lambda_i$ and $\tilde{\zeta}_i$, such that for those in the group $\group = 1, \dots, B$, $\tilde{p}_{2,a_2,i}$ is the same and is denoted by $p_{2,a_2|a_1,\group}$. The DTR means can be estimated by Eq.~(\ref{ipw_1}). The procedure for individualizing stage-1 randomization probabilities in a SMART-EXAM is available in Web Appendix B. 

\subsection{Theoretical properties of the SMART-EXAM}

The following are three theoretical properties of the SMART-EXAM, and the corresponding proofs are provided in Web Appendix C.

\begin{property}
\label{property_1}
When the participants' preferences and predicted treatment effects are not of concern, a SMART-EXAM with capacity $C_{a_2|a_1}$ can be reduced to a SMART with non-individualized randomization probabilities $p^{0}_{2,a_2} = \frac{C_{a_2|a_1}}{ N^{nr}_{a_1}}$, by setting $\Lambda_{i} = \Lambda_{j}$ and $\tilde{\zeta}_{i} = \tilde{\zeta}_{j}$ for all $i$ and $j$ ($j \neq i$) among non-responders to the initial treatment $a_1$. 
\end{property}

\begin{property}
\label{property_2}
Under the three causal assumptions in the Neyman-Rubin causal framework: 1) sequential exchangeability assumption (SEA); 2) consistency assumption (CA); and 3) positivity assumption (PA), which are detailed in Web Appendix C, the IPW estimator for the value of DTR $\bm{d}_j$ in a SMART-EXAM is a consistent estimator of the expected outcome of $\bm{d}_j$, i.e., $\hat{\mu}_{\bm{d}_j} \xrightarrow{p} \mu_{\bm{d}_j}$.
\end{property}

\begin{property}
\label{property_3}
The large-sample distribution of $\hat{\mu}_{\bm{d}_j}$ for DTR $\bm{d}_j$ is $\sqrt{N}(\hat{\mu}_{\bm{d}_j} - \mu_{\bm{d}_j}) \xrightarrow{d} N(0, \sigma^2_{\bm{d}_j})$, where
\begin{equation}
\begin{aligned}
 \sigma^2_{\bm{d}_j}&=\frac{\pi_{1, d_{j1}}}{p_{d_{j1}}}\{ \sigma^2_{s(d_{j1}, 0)}
 +(\mu_{\bm{d}_j}-\mu_{s(d_{j1},0)})^2 \} \\
 &+\sum_{g} \left[ \frac{(1-\pi_{d_{j1}})\pi_{g|d_{j1}}}{p_{1,d_{j1}} p_{2,d_{j2}|a_1,g}} \times \{ \sigma^2_{s(d_{j1},g,d_{j2})}+(\mu_{\bm{d}_j}-\mu_{s(d_{j1},g,d_{j2})})^2 \} \right], 
\end{aligned}
\end{equation}
with $\mu_{s(d_{j1},g,d_{j2})}=E[Y|A_{1}=d_{j1}, R=0,G=g, A_{2}=d_{j2}]$, $\sigma^2_{s(d_{j1},g,d_{j2})}=\Var(Y|A_{1}=d_{j1}, R=0, G=g, A_{2}=d_{j2})$, $\mu_{s(d_{j1},0)}=E[Y|A_{1}=d_{j1}, R=1]$, $\sigma^2_{s(d_{j1},0)}=\Var(Y|A_{1}=d_{j1}, R=1)$, and $\pi_{g|d_{j1}} = \pr(G=g|A_1=d_{j1},R=0)$
\end{property}

\section{Simulations}
\label{sec:section4}

Using 500 simulation replicates, we compare the performance of the SMART-EXAM with the SMART in different settings detailed in \hyperlink{sec:section4.1}{Section 4.1}. 
\subsection{Data generation}
\label{sec:section4.1}

Table~\ref{table_1} provides the parameter specification for generating data from SMART designs. As discussed in \hyperref[sec:section3.1]{Section 3.1} that there may be positive or negative associations between treatment effects and preferences, we consider two simplified scenarios for generating participants' preferences. The first assumes a negative association between preferences $\Lambda_{i}$ and treatment effects $\zeta_{i}$, i.e., participants with higher effects of $A_2=a_2^e$ are less likely to prefer $A_2=a_2^e$ compared to those with lower effects. The second assumes a positive association between preference $\Lambda_{i}$ and treatment effect $\zeta_{i}$. It is important to note that our approach to generating preference indicators does not imply that the SMART-EXAM requires a specific association between these two measures. Rather, it serves as a means to explore the performance of the SMART-EXAM with different degrees of trade-off between preferences and treatment effects. We reemphasize that, since our goal was not to capture the full complexity of participants' preferences, we only assume a simplified association between these two measures in simulation studies, which is sufficient to provide valuable insights. 

Furthermore, we consider different parameter sets for $\epsilon$ and $\eta$ in SMART-EXAMs detailed in Table~\ref{table_1}. Recall that $\epsilon$ controls the range of the randomization probabilities and $\eta$ is the coefficient in the price function. Parameter sets 1) - 3) are to explore the impact of $\epsilon$ on the performance of SMART-EXAMs, and SMART-EXAMs with smaller values of $\epsilon$ are expected to generate more extreme probabilities for improving participants' welfare; while parameter sets 2) and 4) - 6) are to investigate the impact of $\eta$ on the performance of SMART-EXAMs. The predicted treatment effects $\hat{\zeta}_{i}$ are derived based on pilot SMARTs with different sample sizes $N^p=100,200,300$ simulated in the same manner as above. These pilot SMARTs are to investigate the sensitivity of the SMART-EXAM to the quality of the external information. We also consider different sample sizes for the full-scale SMART $N=200,300,400,500$ and treatment capacities $C_{1|a_1}$, including $0.5N^{nr}_{a_1}$, $0.6N^{nr}_{a_1}$, and $0.7N^{nr}_{a_1}$, denoted by $C=0.5$, $C=0.6$, and $C=0.7$ respectively for the simplicity of illustration. Furthermore, we include a SMART-EXAM with adaptive randomization named SMART-AR-EXAM and assume equal treatment capacity, i.e., $C=0.5$. In a SMART-AR-EXAM, the first $N_{\min}=100$ participants will get randomized using balanced randomization. The data from these participants instead of a pilot SMART are then used to estimate treatment effects and to generate the randomization probabilities for the remaining participants. The SMART-AR-EXAM and the SMART-EXAM resemble the duality between adaptive and fixed RCTs, where the former is an adaptive design that allows for updating randomization probabilities based on the accumulating data, while the latter generates fixed but unbalanced randomization probabilities based on external information. 

\begin{table}[htb]
    \centering
    \caption{Parameter specification for generating data from SMART designs in the simulation study. $\zeta_{2,1,i}$ is the treatment effect of $a_2=1$ compared to $a_2=-1$ on the outcome; $\epsilon$ is the parameter that controls the range of the randomization probabilities generated by SMART-EXAMs; $\eta$ is the coefficient for the treatment effect in the price function. }
    \label{table_1}
    \small
    \begin{tabular}{p{5cm} p{8cm}}
    \hline
     Variable & Specification \\
    \hline
     The stage-1 treatment $A_1$ & $(A_1+1)/2 \sim \ber(0.5)$ \\
    Intermediate response status $R$ &  $R \sim \ber(0.5)$ \\
     Tailoring variables $O_{21}$ and $O_{22}$ & Standard normal distribution $N(0,1)$\\
     \multirow{2}{3cm}{The outcome $Y$}  &  For non-responders:  $Y_i=2 - A_{1i} + A_{2i} + 0.5 A_{1i} A_{2i} -0.5 O_{21,i} A_{2i} +0.5 O_{22,i} A_{2i} + \tau, \quad \tau \sim N(0,3^2)$ \\
     & For responders: $Y_i=3+A_{1i} + \tau, \quad \tau \sim N(0,3^2)$ \\
     \multirow{2}{*}{Preference indicator $\Lambda_i$} & Negative association: $\Pr(\Lambda_{i}=1)=\logit^{-1}(-0.2\zeta_{2,1,i}+1 )$ \\
     & Positive association: $\Pr(\Lambda_{i}=1)=\logit^{-1}(0.2\zeta_{2,1,i}+0.5)$ \\
     $\epsilon$ and $\eta$ & 1) $\epsilon=0.1,\eta=-1$; 2) $\epsilon=0.2,\eta=-1$; 3) $\epsilon=0.3,\eta=-1$; 4) $\epsilon=0.2,\eta=-0.7$; 5) $\epsilon=0.2,\eta=-0.4$; 6) $\epsilon=0.2,\eta=-0.1$ \\
    \hline
    \end{tabular}
\end{table}

\subsection{Metrics for the design performance}
\label{sec:section4.2}

The true value of $\mu_{\bm{d}_j}$ is approximated via IPW based on a simulated SMART of size $100,000$. The DTR mean $\hat{\mu}_{\bm{d}_j}$ and its variance $\hat{\sigma}_{\bm{d}_j}^2$ are estimated by Eq.~(\ref{ipw_1}) and Eq.~(\ref{ipw_3}), respectively. The metrics to compare these designs can be divided into two categories: the first is the welfare of enrolled participants, including the participants' average outcome $\bar{Y}=\frac{1}{N^{nr}}\sum_{i=1}^{N^{nr}} Y_{i}$, where $N^{nr}$ is the number of non-responders, the average probability of receiving the preferred treatment $\bar{u}=\frac{1}{N^{nr}} \sum_{i=1}^{N^{nr}} u_i$, and the number of participants in DTR $\bm{d}_j$ denoted by $N_{\bm{d}_j}$. Note that we only focus on non-responders when comparing participants' welfare because responders only have one treatment option at stage 2. The second is the learning ability of SMART designs, evaluated by whether they correctly estimate the true optimal DTR or not denoted by $I(\hat{\bm{d}}^{\ast} = \bm{d}^{\ast})$, and the true value of the estimated optimal DTR $\mu_{\hat{\bm{d}}^{\ast}}$, which can be interpreted as the expected outcome if the estimated optimal DTR $\hat{\bm{d}}^{\ast}$ learned from the current SMART is given to the entire population. These metrics are collected in each simulation replicate and averaged over to get the expected value. 

\subsection{Simulation results}
\label{sec:section4.3}

\begin{figure}[htb]
\centering
\includegraphics[width=0.8\linewidth]{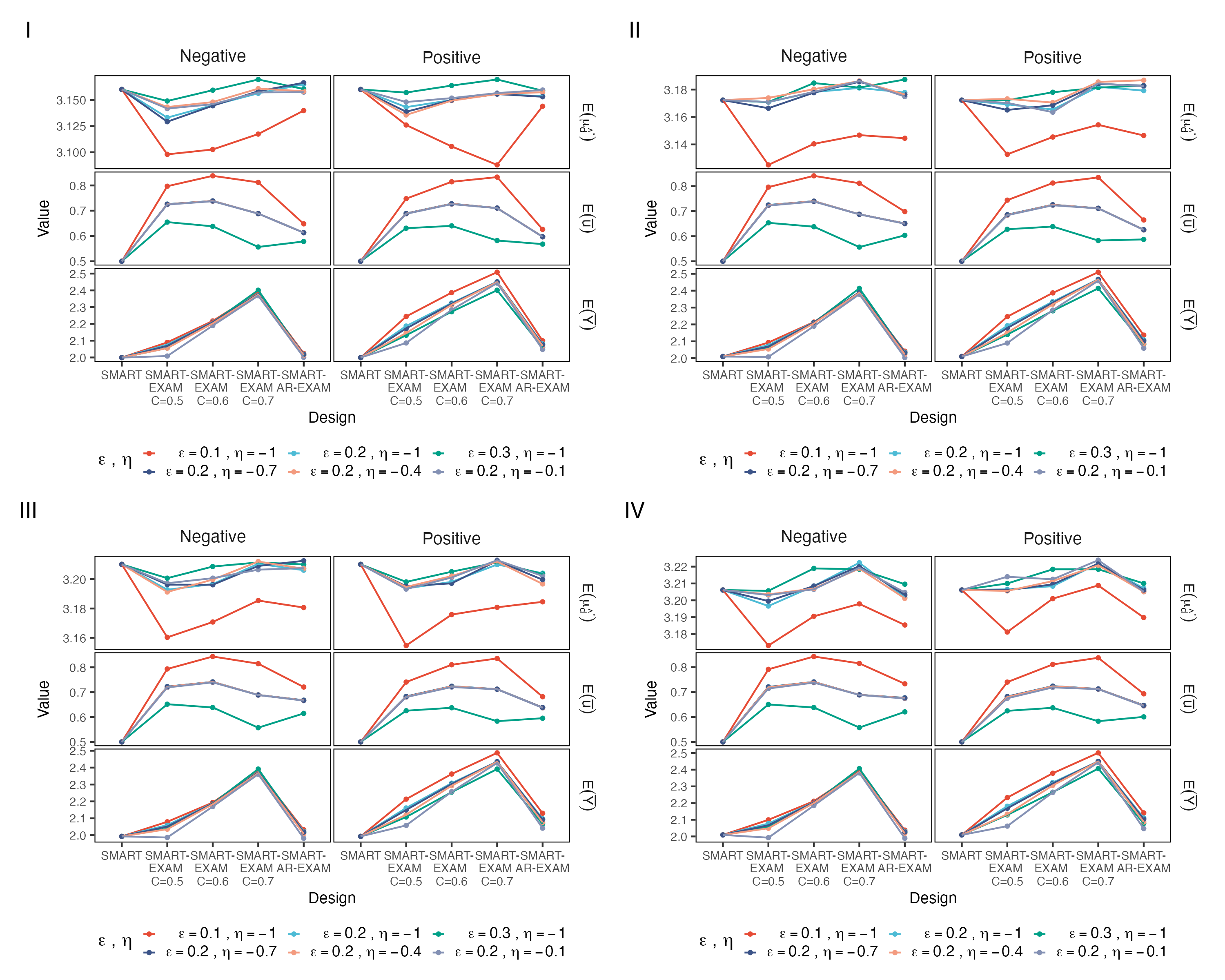}
\caption{Simulation results for the setting with a pilot SMART of size $N^p=200$. I, II, III, and IV correspond to the sample size $N=200, 300, 400, 500$ respectively. The left (right) panel of each subfigure corresponds to the negative (positive) association between preferences and treatment effects. $E[\mu_{\hat{\bm{d}}^{\ast}}]$ denotes the Monte Carlo estimate of the true value of the estimated optimal DTR; $E[\bar{u}]$ denotes the Monte Carlo estimate of the probability of being assigned to the preferred treatment, and $E[\bar{Y}]$ denotes the Monte Carlo estimate of the mean outcome. }
\label{simu_plot_1}
\end{figure}

This subsection gives the simulation results for the setting with a pilot SMART of size $N^p=200$. Results for other settings with pilot SMARTs of size $N^p=100,300$ are provided in Web Appendix D. Figure~\ref{simu_plot_1} shows that, when setting equal treatment capacity for each treatment, the SMART has better learning ability represented by $E[\mu_{\hat{\bm{d}}^{\ast}}]$, i.e., the Monte Carlo estimate of the true value of the estimated optimal DTR, compared to the SMART-EXAM $C=0.5$, except for the cases with sample size $N=500$ and a positive association between treatment effects and preferences, where SMART-EXAM with $C=0.5$ performs slightly better than the SMART. However, as $\epsilon$ increases, the learning abilities of SMART-EXAM with $C=0.5$ become closer to the SMART. A possible explanation might be that a lower value of $\epsilon$ corresponds to more extreme randomization probabilities, resulting in higher variances of estimated DTR means and thus less accurate estimation of optimal DTRs. As the sample size $N$ increases, the learning abilities for all these designs improve.

In addition, all the SMART-EXAM designs demonstrate a substantial improvement in the average probability of receiving the preferred treatment $E[\bar{u}]$ and final outcome $E[\bar{Y}]$ compared to the SMART, which is as expected given that the SMART does not take into account the participants' preferences and treatment effects during randomization. While the SMART-AR-EXAM performs better in selecting the true optimal DTR compared to the SMART-EXAM with $C=0.5$ in most scenarios, it offers fewer improvements in terms of $E[\bar{u}]$ and $E[\bar{Y}]$. This is reasonable given that some participants in a SMART-AR-EXAM are randomized using balanced randomization probabilities instead of individualized randomization based on treatment effects and preferences. 

One can also see from Figure~\ref{simu_plot_1} that when $C$ increases from $0.5$ to $0.7$,  $E[\mu_{\hat{d}^{\ast}}]$ and $E[\bar{Y}]$ increase, as more participants are randomized to the true superior treatment $A_2=1$. However, in reality, the true superior treatment is unknown and the treatment capacity $C$ should be specified based on various factors mentioned in \hyperref[sec:section3.1]{Section 3.1}. The coefficient $\eta$ has a relatively smaller impact on the performance of SMART-EXAMs. Nonetheless, as the absolute value of $\eta$ increases,  there is a slight upward trend in the mean outcomes $E[\bar{Y}]$. This finding aligns with the underlying concept of $\eta$, where higher values of $\eta$ amplify the predicted treatment effect's impact on the randomization probabilities, resulting in better performance in improving the outcome. Moreover, SMART-EXAMs demonstrate higher mean outcomes $E[\bar{Y}]$ in scenarios with a positive association between treatment effects and preferences compared to those with a negative association, which is not unanticipated given that the tradeoff between preferences and treatment effects of a positive association is expected to be less than that of a negative association. We also provide results under settings with other pilot SMARTs in Web Figures 1-2 in Web Appendix D, which show similar patterns to the aforementioned findings.   

In addition to IPW estimators, we explore using AIPW for estimating DTR means. Furthermore, we employ other metrics to evaluate the operating characteristics of the SMART-EXAM designs. More specifically, we consider 1) the probabilities of being selected as the optimal DTR for each DTR $\bm{d}_j$ by IPW and AIPW, denoted as $\Pr(\bm{d}_j=\hat{\bm{d}}^{\ast IPW})$ and $\Pr(\bm{d}_j=\hat{\bm{d}}^{\ast AIPW})$; 2) the average estimated value of DTR $\bm{d}_j$ by IPW and AIPW, denoted as $E[\hat{\mu}^{IPW}_{\bm{d}_j}]$ and $E[\hat{\mu}^{AIPW}_{\bm{d}_j}]$, and 3) the average number of participants assigned to DTR $\bm{d}_j$, i.e., $\bar{N}_{\bm{d}_j}$ for the scenario with sample size $N=200$, $N^p=200$, $\eta=-1$ and $\epsilon=0.1, 0.2, 0.3$. Web Tables 1-3 in Web Appendix D show that, in all these SMART designs, $E[\hat{\mu}^{IPW}_{\bm{d}_j}]$ and $E[\hat{\mu}^{AIPW}_{\bm{d}_j}]$ are close to the true value of the corresponding DTR; thus, all of these designs achieve the desired performance of estimating DTR means. Furthermore, when $\epsilon=0.3$, all SMART-EXAMs exhibit better performance than the SMART in terms of selecting the true optimal DTR, except that the SMART-EXAM with $C=0.5$ performs slightly worse than the SMART when there is a negative association between treatment effects and preferences; the probability of selecting the true optimal DTR for the SMART-EXAM with $C=0.5$ is $0.812$, whereas for the SMART it is $0.824$. The SMART-AR-EXAM demonstrates a comparable performance to the SMART in terms of selecting optimal DTRs and estimating DTR means, thus it may serve as an alternative design that can incorporate participants' welfare when there is no prior information about treatment effects. 

\section{Empirical Application}
\label{sec:section5}

In this section, we illustrate the merits of the SMART-EXAM using the aforementioned SMART ADHD study as a previous/pilot SMART. Investigators from the University of Michigan provide simulated data based on the real data from this two-stage ADHD SMART with a sample size $N=150$, comprised of 51 responders and 99 non-responders \citep{almirall2023adhd}. The baseline covariates include $O_{11}$: the indicator for oppositional defiant disorder (ODD) diagnosis, coded as 1/0, $O_{12}$: continuous ADHD score reflecting ADHD symptoms at the end of the previous school year, $O_{13}$: the indicator for medication prior to first-stage treatments, coded as 1/0, and $O_{14}$: the indicator for whether the race is White or not, coded as 1/0. The intermediate potential tailoring variables include $O_{21}$: the number of months until non-response and $O_{22}$: the indicator for adherence to the stage-1 treatments, coded as 1/0 for higher and lower adherence. Imagine that researchers wish to further explore the effectiveness of the embedded DTRs for children with ADHD, and plan to conduct a new SMART. To assess the potential benefits in terms of participants' welfare gained by implementing a SMART-EXAM compared to a conventional SMART, we simulate data from these designs based on the original data. Note that for the illustration purpose, we assume the intermediate decision point is at one fixed time point for the SMART ADHD study thus ignoring the variable $O_{21}$. The specification for the required parameters is provided in Web Table 4 in Web Appendix D. 

We assume equal treatment capacities, i.e., $C=0.5$ for the SMART and SMART-EXAMs. For SMART-EXAMs, we specify $\eta=-1$, $m=1$, and consider various values of $\epsilon=0.1,0.2,0.3$. As stated in \hyperref[sec:section3.1]{Section 3.1}, even though the original SMART ADHD study did not collect participants' preferences for stage-2 treatments, it is reasonable to assume that, participants with lower adherence to the initial treatment, i.e., $O_{22}=0$, are more likely to prefer $A_2=1$, i.e., augmentation, than those with higher adherence, i.e., $O_{22}=1$. This corresponds to the scenario with a positive association between treatment effects and preferences in \hyperlink{sec:section4.1}{Section 4.1}. We consider three possible settings for the preference data based on the above assumption, which is detailed in Web Table 4 in Web Appendix D.  

\subsection{Application results}
\label{sec:section5.2}

As shown in Table~\ref{table_3}, the SMART-EXAM with $\epsilon=0.1$ performs worse than the other designs in terms of selecting the true optimal DTR, denoted by $\pr(\hat{\bm{d}}^{\ast IPW} = \bm{d}^{\ast})$, in all these scenarios. However, it outperforms the other designs in terms of improving the probability of receiving the preferred treatment and the mean outcomes. As $\epsilon$ increases, the learning ability of the SMART-EXAM improves and becomes closer to that of the SMART, while still continuing to improve $E[\bar{u}]$ and $E[\bar{Y}]$ compared to the SMART. These findings convey that, based on the original data, when conducting a new SMART for children with ADHD, the SMART-EXAM can increase the chance of allocating participants to their preferred treatment and improve the outcomes for enrolled participants, while still maintaining its effectiveness in estimating optimal DTRs with a moderate value of $\epsilon$.

\begin{table}
\centering
\caption{The application results. $\pr(\hat{\bm{d}}^{\ast IPW} = \bm{d}^{\ast})$ is the probability of selecting the true optimal DTR by IPW; $E[\mu_{\hat{\bm{d}}^{\ast IPW}}]$ is the Monte Carlo estimate of the true value of the estimated optimal DTR by IPW; $E[\bar{u}]$ is the Monte Carlo estimate of the probability of receiving the preferred treatment; $E[\bar{Y}]$ is the Monte Carlo estimate of the mean outcome among non-responders. The true optimal DTR and the true DTR means are approximated using a simulated SMART of sample size 10,000. }
\label{table_3}
\small
\begin{tabular}{llcccccccc}
  \hline
Scenario& Design & $\pr(\hat{\bm{d}}^{\ast IPW} = \bm{d}^{\ast})$ & $E[\mu_{\hat{\bm{d}}^{\ast IPW}}]$ & $E[\bar{u}]$ & $E[\bar{Y}]$ \\
 \hline
S1 & SMART & 0.984 & 3.504 & 0.500 & 3.002 \\ 
   & SMART-EXAM $\epsilon=0.1$ & 0.922 & 3.468 & 0.817 & 3.263 \\ 
   & SMART-EXAM $\epsilon=0.2$ & 0.978 & 3.500 & 0.741 & 3.200 \\ 
   & SMART-EXAM $\epsilon=0.3$ & 0.984 & 3.504 & 0.664 & 3.136 \\ 
  S2 & SMART & 0.984 & 3.504 & 0.500 & 3.002 \\ 
   & SMART-EXAM $\epsilon=0.1$ & 0.942 & 3.480 & 0.792 & 3.156 \\ 
   & SMART-EXAM $\epsilon=0.2$ & 0.960 & 3.491 & 0.722 & 3.115 \\ 
   & SMART-EXAM $\epsilon=0.3$ & 0.980 & 3.502 & 0.652 & 3.074 \\ 
  S3 & SMART & 0.984 & 3.504 & 0.500 & 3.002 \\ 
   & SMART-EXAM $\epsilon=0.1$ & 0.940 & 3.479 & 0.745 & 3.245 \\ 
   & SMART-EXAM $\epsilon=0.2$ & 0.980 & 3.501 & 0.687 & 3.186 \\ 
   & SMART-EXAM $\epsilon=0.3$ & 0.988 & 3.506 & 0.629 & 3.128 \\ \hline
\end{tabular}
\end{table}

\section{Discussion}
\label{sec:section6}

A conventional SMART randomizes participants with non-individualized and balanced randomization probabilities. Despite its implementation simplicity and desirable performance in comparing embedded DTRs, it faces some ethical issues. To mitigate these issues, we propose the SMART-EXAM, a novel SMART design that incorporates participants' preferences and predicted treatment effects into the randomization procedure, to advance health promotion among both the participants enrolled in the trial and future patients. We provide a detailed illustration of conducting a SMART-EXAM and assess its empirical properties through an extensive simulation study. The simulation results demonstrate that the SMART-EXAM can improve the average probability of receiving the preferred treatment and the final outcome compared with the conventional SMART, while also achieving a comparable ability to construct optimal DTRs for future patients with a moderate value of $\epsilon$, which controls the range of randomization probabilities. We argue that by incorporating participant preferences and predicted treatment effects, the SMART-EXAM has the potential to encourage more people to participate in the trial and alleviate the lost-to-follow-up issue, finally resulting in a more diverse and representative sample compared to the conventional SMART. 

It is important to note that each SMART design has its own strengths and weaknesses. The choice of design should depend on the primary research goals and specific contexts. For instance, if patient enrollment is challenging due to their reluctance to get randomized with balanced probabilities, the SMART-EXAM may be a better choice. Conversely, if the primary focus is on improving treatment effectiveness for enrolled participants, the SMART-AR may be more appropriate, as it continuously updates randomization probabilities based on accumulating information about treatment effects. When maximizing statistical power becomes a priority, the conventional SMART may be a suitable choice. The SMART-EXAM offers investigators the flexibility to adjust experimental parameters based on their specific research focus and clinical experience. Specifically, when the primary focus is to maximize the statistical power for future patients instead of participants' welfare, the SMART-EXAM can be \textit{reduced} to a SMART with non-individualized randomizations by setting $\Lambda_i=\Lambda_j$ and $\zeta_i=\zeta_j, j \neq i$ for all participants. In addition, higher values of $\epsilon$ can also increase the learning ability of the SMART-EXAM. In contrast to many adaptive trials that lack strict capacity constraints, the SMART-EXAM incorporates a pre-specified treatment capacity $C_{a_2|a_1}$ to control the expected number of participants in each treatment. This feature resembles the SMART where the sample size is predetermined for each treatment arm and is particularly useful when treatment resources are limited or when investigators wish to prespecify the sample size for each treatment for the purpose of grant application. 

The SMART-EXAM can be potentially extended to SMARTs with more than two stages or more than two treatment options at each stage. Furthermore, \citet{walter2017beyond} summarized a statistical framework for estimating treatment effects, selection effects, and preference effects based on data from patient-preference designs introduced in \hyperref[sec:section1]{Section 1}. The SMART-EXAM serves as another potential design to quantify these estimands.

Despite its great potential, several limitations of the SMART-EXAM need to be acknowledged. One limitation of the SMART-EXAM, which is also shared by various types of adaptive designs \citep{david2023response}, is that there are no explicit sample size calculation formulae for these designs. Simulation studies based on postulated parameters are needed to ensure sufficient statistical power. Given a sufficient sample size, the SMART-EXAM is expected to effectively identify the optimal DTR due to the consistency of the IPW estimators of DTR means, i.e., as the sample size increases, the estimated DTR means all converge to their true values, enabling the accurate identification of the true optimal DTR. Additionally, participants in a SMART-EXAM may feel disappointed when receiving a treatment they dislike after they have indicated their preferences. To manage expectations and minimize potential disappointment, it is crucial to ensure that participants have a realistic understanding of the SMART-EXAM by emphasizing that while this design could improve the average probability of receiving the preferred treatment, it cannot guarantee that every participant will be allocated to their preferred treatment. Another potential solution is to predict the preference using previous data resources instead of letting participants indicate their preferences. An additional challenge in a SMART-EXAM is the existence of participants without specific treatment preferences. One potential approach is to use balanced probabilities for participants without preferences, a straightforward adjustment that can be integrated into the existing SMART-EXAM algorithm. In addition, researchers may consider combining the no-preference group with one of the preference groups based on the specific context. For example, \citet{chakraborty2013inference} merged participants who had no preferences with those who preferred augmentation with a new treatment. This combined group can be considered as participants open to either treatment option. 


\section*{ACKNOWLEDGMENTS}

We would like to thank Haruka Kiyohara, the referees, the associate editor, and the co-editor for providing insightful comments and suggestions to improve the paper. B. Chakraborty would like to acknowledge support from an Academic Research Fund Tier 2 grant (MOE-T2EP20122-0013) from the Ministry of Education, Singapore, as well as the start-up grant from the Duke-NUS Medical School, Singapore.

\section*{DATA AVAILABILITY STATEMENT}
The data used in this paper is publicly available at \url{http://www-personal.umich.edu/~dalmiral/atsworkshops.html}.

\bibliography{arxiv_v1}

\end{document}


\title{Supporting Information for ``Incorporating Participants' Welfare into Sequential Multiple Assignment Randomized Trials''\vspace{2cm}}

\author[1]{Xinru Wang}
\author[2, 3]{Nina Deliu}
\author[4]{Yusuke Narita}
\author[1, 5, 6, 7]{Bibhas Chakraborty}

\footnotesize{

\affil[1]{Centre for Quantitative Medicine, Duke-NUS Medical School, Singapore} 
\affil[2]{MRC - Biostatistics Unit, University of Cambridge, Cambridge, UK 
} 
\affil[3]{MEMOTEF Department, Sapienza University of Rome, Rome, Italy} 
\affil[4]{Department of Economics and Cowles Foundation, Yale University, New Haven, Connecticut, USA} 
\affil[5]{Program in Health Services and Systems Research, Duke-NUS Medical School, Singapore}
\affil[6]{Department of Statistics and Data Science, National University of Singapore, Singapore}
\affil[7]{Department of Biostatistics and Bioinformatics, Duke University, Durham, North Carolina, USA}
}
\date{}

\maketitle

\section*{Web Appendix A: The Algorithm for the SMART-EXAM}

\begin{algorithm}[H]
\begin{spacing}{1}
  \caption{}
  \label{AIPal}
  \begin{algorithmic}[1]
   \INPUT $\{C_{a_2|a_1}\}_{a_2=1,-1}$: the capacity for the stage-2 treatments among non-responders to treatment $a_1$;  $\{\Lambda_i\}_{i=1,\dots,N_{a_1}^{nr}}$: the vector of preference indicators for the non-responders to $a_1$; $m=1$: the budget for ``buying'' randomization probabilities for treatment $A_2=a_2^e$; $\{\tilde{\zeta}_i\}_{i=1,\dots,N_{a_1}^{nr}}$: the vector of categorized estimated effects of treatment $a_2^e$ over treatment $a_2^o$ on the outcome for the non-responders to $a_1$; $\eta \in (-1,0)$: the coefficient of treatment effects in the price function; $\kappa$: the threshold for control the sum of the square differences between the expected number of participants and the capacity for each treatment; $M$: the threshold for iteration times for updating the value of $\beta$; $l$: the step size for updating the value of $\beta$.
    \OUTPUT $\{p_{2,a_2^e,i}\}_{i=1,\dots,N_{a_1}^{nr}}$: the randomization probabilities for $a_2^e$ among the non-responders to $a_1$; 
    \Function{INITBETA}{$\{\tilde{\zeta}_i\}_{i=1,\dots,N_{a_1}^{nr}}$} \Comment{Set the initial value of $\beta$}
    \State $\beta = -\underset{i \in 1,\dots,N_{a_1}^{nr}}{\max} |\tilde{\zeta}_i|$
    
    \Return{$\beta$}
    \EndFunction   
    
    \Function{PRICE}{$\eta$, $\beta$, $\{\tilde{\zeta}_i\}_{i=1,\dots,N_{a_1}^{nr}}$}  \Comment{Get the price of a unit randomization probability for $a_2^e$}
    \For{$i \in 1, \dots, N^{nr}_{a_1}$}
    \State $\Psi_i = \eta \tilde{\zeta}_i + \beta$
    \EndFor
    
    \Return{$\{\Psi_i\}_{i=1,\dots,N_{a_1}^{nr}}$}
    \EndFunction
    
    \Function{DEMAND}{$\{\Psi_i\}_{1,\dots,N_{a_1}^{nr}}$,$m=1$, $\{\Lambda_i\}_{1,\dots,N_{a_1}^{nr}}$,$\epsilon$}   \Comment{Solve the utility maximization function subject to the budget constraint}
    \For{$i \in 1, \dots, N^{nr}_{a_1}$}
    \State $\hat{p}_{2,a_2^e,i} \longleftarrow \underset{ p_{2,a_2^e,i} \in \mathcal{P}}{\argmax} \ u_i \quad s.t. \ p_{2,a_2^e,i}\ \Psi_{i}  \leq m$  \Comment{$u_i=p_{2,1,i}\Lambda_i+(1-p_{2,1,i})(1-\Lambda_i)$}
    \EndFor
    \State  Find the minimum number $q \in (0,1)$ that satisfies $\tilde{p}_{2, a_2^e,i} \in [\epsilon,1-\epsilon]$ for $i=1,\dots,N_{a_1}^{nr}$, where $\tilde{p}_{2, a_2^e,i} = (1-q) \hat{p}_{2, a_2^e,i} + q p^0_{2, a_2^e}$.
    
    \Return{$\{\tilde{p}_{2, a_2^e,i}\}_{i=1,\dots,N_{a_1}^{nr}}$}
    \EndFunction
    
     \Function{EXCESSDEMAND}{$\{\tilde{p}_{2,a_2,i}\}_{1,\dots,N_{a_1}^{nr}}$}   \Comment{Get the excess demand for stage-2 treatment}
    \For{$a_2 \in \{ 1, -1\}$}
    \State $d_{a_2} \longleftarrow \sum_i^{N^{nr}_{a_1}} \tilde{p}_{2,a_2,i} - C_{a_2|a_1}$
    \EndFor
    
    \Return{$\{d_{a_2}\}_{a_2 = 1, -1}$}
    \EndFunction
    
         \Function{CLEARERROR}{$\{d_{a_2}\}_{a_2=1,-1}$}. \Comment{Get the market clear error}

   \State error $\longleftarrow \sum_{a_2}d_{a_2}^2/\sum_{a_2}C_{a_2|a_1}$ 
    
    \Return{error}
    \EndFunction

    \Function{BETANEW}{$\beta,d_{a_2^e},l$}   \Comment{Adjust $\beta$ to set new prices}
    \State $\beta^{new} \longleftarrow \beta+d_{a_2^e}/{l}$ 
    
    \Return{$\beta^{new}$}
    \EndFunction
    
        \algstore{myalg}
  \end{algorithmic}
  \end{spacing}
\end{algorithm}

\clearpage

\begin{algorithm}

\ContinuedFloat
  \caption{ (continued)}
  \begin{algorithmic}[1]  
  \algrestore{myalg}  

\Function{MAIN}{$\{\Psi_i\}_{1,\dots,N_{a_1}^{nr}}$, $\eta$, $m$, $\{\Lambda_i\}_{1,\dots,N_{a_1}^{nr}}$, $\kappa$, $l$, $\epsilon$}   
 \State $r \longleftarrow$ 1   
  \State $\beta \longleftarrow$ INITBETA($\{\tilde{\zeta}_i\}_{1,\dots,N_{a_1}^{nr}}$) 
  \State  $\{\Psi_i\}_{1,\dots,N_{a_1}^{nr}} \longleftarrow$ PRICE($\eta$, $\beta$, $\{\tilde{\zeta}_i\}_{1,\dots,N_{a_1}^{nr}}$) 
  \State  $\{\tilde{p}_{2,a_2^e,i}\}_{1,\dots,N_{a_1}^{nr}} \longleftarrow$ DEMAND($\{\Psi_i\}_{1,\dots,N_{a_1}^{nr}}$,$m=1$, $\{\Lambda_i\}_{1,\dots,N_{a_1}^{nr}}$,$\epsilon$) 
    
  \State  $\{d_{a_2}\}_{a_2=1,-1} \longleftarrow$ EXCESSDEMAND($\{\tilde{p}_{2,a_2,i}\}_{1,\dots,N_{a_1}^{nr}}$) 
  \State error $\longleftarrow$ CLEARERROR($\{d_{a_2}\}_{a_2=1,-1}$)

  \If{error $\leq \kappa$}
 \State \textbf{Break}
\EndIf

  \While{True}
  \If{$r > M$}
    \State $r \longleftarrow$ 1
  \State $\kappa \longleftarrow \kappa + 0.002$
  \State $\beta \longleftarrow$   INITBETA($\{\tilde{\zeta}_i\}_{1,\dots,N_{a_1}^{nr}}$) 
  \State  $\{\Psi_i\}_{1,\dots,N_{a_1}^{nr}} \longleftarrow$ PRICE($\eta$, $\beta$, $\{\tilde{\zeta}_i\}_{1,\dots,N_{a_1}^{nr}}$) 
  \State  $\{\tilde{p}_{2,a_2^e,i}\}_{1,\dots,N_{a_1}^{nr}} \longleftarrow$ DEMAND($\{\Psi_i\}_{1,\dots,N_{a_1}^{nr}}$,$m=1$, $\{\Lambda_i\}_{1,\dots,N_{a_1}^{nr}}$,$\epsilon$) 
    
  \State  $\{d_{a_2}\}_{a_2=1,-1} \longleftarrow$ EXCESSDEMAND($\{\tilde{p}_{2,a_2,i}\}_{1,\dots,N_{a_1}^{nr}}$) 
  \State error $\longleftarrow$ CLEARERROR($\{d_{a_2}\}_{a_2=1,-1}$)
  \Else
  \State $r \longleftarrow r + 1$
  \State $\beta \longleftarrow$ BETANEW($\beta,d_{a_2^e},l$)
  \State  $\{\Psi_i\}_{1,\dots,N_{a_1}^{nr}} \longleftarrow$ PRICE($\eta$, $\beta$, $\{\tilde{\zeta}_i\}_{1,\dots,N_{a_1}^{nr}}$) 
  \State  $\{\tilde{p}_{2,a_2^e,i}\}_{1,\dots,N_{a_1}^{nr}} \longleftarrow$ DEMAND($\{\Psi_i\}_{1,\dots,N_{a_1}^{nr}}$,$m=1$, $\{\Lambda_i\}_{1,\dots,N_{a_1}^{nr}}$,$\epsilon$) 
    
  \State  $\{d_{a_2}\}_{a_2=1,-1} \longleftarrow$ EXCESSDEMAND($\{\tilde{p}_{2,a_2,i}\}_{1,\dots,N_{a_1}^{nr}}$) 
  \State error $\longleftarrow$ CLEARERROR($\{d_{a_2}\}_{a_2=1,-1}$)
  \EndIf
   \If{error $\leq \kappa$}
 \State \textbf{Break}
\EndIf
  \EndWhile
  \State  \Return{$\{\tilde{p}_{2,a_2^e,i}\}_{1,\dots,N_{a_1}^{nr}}$} 
    \EndFunction
    \end{algorithmic}
\end{algorithm}

\clearpage

\section*{Web Appendix B: A two-stage two-treatment SMART-EXAM with the randomization probabilities individualized at both stages.}
While there are various advanced statistical techniques that can be utilized to estimate treatment effects, e.g., random forest, boosting, and regression trees, here we restrict our focus to Q-learning with linear regression for the simplicity of illustration. Define the stage-2 Q-function for non-responders as
\begin{equation}
\label{appendix_1}
Q_2(\bm{H}_{2i}, A_{2i}; \bm{\gamma}_2, \bm{\alpha}_2) = \bm{\gamma}_2^{T} \bm{H}_{20i}+ \bm{\alpha}_2^{T}\bm{H}_{21i} A_{2i},
\end{equation}
where $\bm{H}_{20i}$ and $\bm{H}_{21i}$ are different features of $\bm{H}_{2i}$ and $\bm{H}_{21i}$ is the vector of covariates in $\bm{H}_{2i}$ that are believed to interact with the treatment. $\bm{\gamma}_2$ and $\bm{\alpha}_2$ are estimated through the Ordinary Least Square method (OLS):
\begin{equation}
\label{appendix_2}
(\hat{\bm{\gamma}}_2, \hat{\bm{\alpha}}_2) = \underset{\bm{\gamma}_2, \bm{\alpha}_2}{\argmin} \sum^{N^{nr}}_{i=1} (Y_i - Q_2(\bm{H}_{2i}, A_{2i}; \bm{\gamma}_2, \bm{\alpha}_2))^2,
\end{equation}
where $N^{nr}$ denotes the number of non-responders to the initial treatment.

Define the stage-1 Q-function as
\begin{equation}
\label{appendix_3}
Q_1(\bm{H}_{1i}, A_{1i}; \bm{\gamma}_1, \bm{\alpha}_1) = \bm{\gamma}_1^{T} \bm{H}_{10i}+ \bm{\alpha}_1^{T}\bm{H}_{11i} A_{1i}.
\end{equation}
$\bm{\gamma}_1, \bm{\alpha}_1$ can be estimated by:
\begin{equation}
\label{appendix_4}
(\hat{\bm{\gamma}}_1, \hat{\bm{\alpha}}_1) = \underset{\bm{\gamma}_1, \bm{\alpha}_1}{\argmin} \sum_{i=1}^{N} (\tilde{Y}_i - Q_1(\bm{H}_{1i}, A_{1i}; \bm{\gamma}_1, \bm{\alpha}_1))^2,
\end{equation}
where $\tilde{Y}_i= \underset{a_{2} \in \{-1,1 \}}{\max} Q_2(\bm{H}_2, a_2; \hat{\bm{\gamma}}_2, \hat{\bm{\alpha}}_2)$ for non-responders and $\tilde{Y}_i=Y_i$ for responders.

Upon enrollment, the researchers collect the baseline covariates $\bm{O}_1$, e.g., age, gender, and race. The parameters $\bm{\gamma}_1$ and $\bm{\alpha}_1$ in Formula~(\ref{appendix_3}) are estimated using Formula~(\ref{appendix_4}) based on the pilot/previous SMART or observational studies, which are then used in combination with the collected patient history data $\bm{O}_1$ to predict the stage-1 individualized treatment effect $\hat{\zeta}_{1, a_1^e, i}$, which equals $2 \hat{\bm{\alpha}}^{T}_1 \bm{H}_{11i}$ when $a_1^e=1$ or equals $-2 \hat{\bm{\alpha}}^{T}_1 \bm{H}_{11i}$ when $a_1^e=-1$. With the predicted treatment effects and elicited preference data, generating randomization probabilities for stage-1 treatments is the same as the procedure described in Section 3.2 of the main paper.
\newpage
\section*{Web Appendix C: Theoretical Properties of SMART-EXAMs and the Corresponding Proofs}

\begin{property}
\label{property_1}
When the participants' preferences and predicted treatment effects are not of concern, a SMART-EXAM with capacity $C_{a_2|a_1}$ can be reduced to a SMART with non-individualized randomization probabilities $p^{0}_{2,a_2} = \frac{C_{a_2|a_1}}{ N^{nr}_{a_1}}$, by setting $\Lambda_{i} = \Lambda_{j}$ and $\tilde{\zeta}_{i} = \tilde{\zeta}_{j}$ for all $i$ and $j$ ($j \neq i$) among non-responders to the initial treatment $a_1$. 
\end{property}

\begin{proof}
Suppose on the contrary that there are some $\tilde{p}_{2,1,i}$ such that $\tilde{p}_{2,1,i} \neq p^0_{2,1}$ when setting $\Lambda_{i} = \Lambda_{j}$ and $\tilde{\zeta}_{i} = \tilde{\zeta}_{j}$ for all $i$ and $j$ ($j \neq i$). Given the condition that $\tilde{\zeta}_{i} = \tilde{\zeta}_{j}$ for all $i$ and $j$ ($j \neq i$), the treatment prices for all participants are the same. With $\Lambda_{i} = \Lambda_{j}$, the utility for each participant is the same, so $\tilde{p}_{2,1,i}=\tilde{p}_{2,1,j} \neq p^0_{2,1}$. 

If $\tilde{p}_{2,1,i}=\tilde{p}_{2,1,j} > p^0_{2,1}= \frac{C_{1|a_1}}{N^{nr}_{a_1}}$, then $\sum_i^{N^{nr}_{a_1}} \tilde{p}_{2,1,i} > \frac{N^{nr}_{a_1} C_{1|a_1}}{N_{a_1}}=C_{1|a_1}$, which is a contradiction to the capacity constraint $\sum_{i}^{N^{nr}_{a_1}}  p_{2,1,i} = C_{1|a_1}$.

If $\tilde{p}_{2,1,i}=\tilde{p}_{2,1,j} < p^0_{2,1}= \frac{C_{1|a_1}}{N^{nr}_{a_1}}$, then $\tilde{p}_{2,-1,i}=\tilde{p}_{2,-1,j} > p^0_{2,-1}= \frac{C_{-1|a_1}}{N^{nr}_{a_1}}$. We can get $\sum_i^{N^{nr}_{a_1}} \tilde{p}_{2,-1,i} >  \frac{N^{nr}_{a_1} C_{-1|a_1}}{N^{nr}_{a_1}}=C_{-1|a_1}$, which is a contradiction to the capacity constraint $\sum_{i}^{N^{nr}_{a_1}}  p_{2,-1,i} = C_{-1|a_1}$.

Given above, we can get $\tilde{p}_{2,1,i}= p^0_{2,1}$ for all non-responders to the initial treatment $a_1$ when setting $\Lambda_{i} = \Lambda_{j}$ and $\tilde{\zeta}_{i} = \tilde{\zeta}_{j}$ for all $i$ and $j$ ($j \neq i$), i.e., the SMART-EXAM is the same as the SMART with non-individualized stage-2 randomization probabilities $p^{0}_{2,a_2} = \frac{C_{a_2|a_1}}{N^{nr}_{a_1}}$.

\end{proof}

\begin{property}
\label{property_2}
Under the three causal assumptions in the Neyman-Rubin causal framework: 1) sequential exchangeability assumption (SEA); 2) consistency assumption (CA); and 3) positivity assumption (PA), which are detailed later, the IPW estimator for the value of DTR $\bm{d}_j$ in a SMART-EXAM is a consistent estimator of the expected outcome of $\bm{d}_j$, i.e., $\hat{\mu}_{\bm{d}_j} \xrightarrow{p} \mu_{\bm{d}_j}$.
\end{property}

To facilitate the proof of the consistent estimator and its asymptotic normality property, we introduce three assumptions under the Neyman-Rubin causal framework:

\begin{itemize}
    \item Sequential exchangeability assumption (SEA):
    
    Under SEA, the treatment allocation at each stage is independent of the potential outcomes conditional on the historical data. In a two-stage SMART-EXAM, $R^{A_1} \perp A_1 | \bm{H}_1$,  $Y^{\bm{d}_j} \perp A_1 | \bm{H}_1$, and $Y^{\bm{d}_j}  \perp A_2 |\bm{H}_2$.
    \item Consistency assumption (CA):
    
    The potential outcome under the observed treatment is the outcome that is actually observed for an individual. In a two-stage SMART-EXAM, $Y^{\bm{d}_j}=Y$ if the participant is randomized to DTR $\bm{d}_j=(d_{j1},d_{j2})$; $R^{a_1}=R$ if $A_1=a_1$.
    \item Positivity assumption (PA):
    
    The probability of receiving each available treatment at each stage should be positive and less than 1.
\end{itemize}

\begin{proof}
    Under the Consistency assumption (CA), the Sequential exchangeability assumption (SEA), and the Positivity assumption (PA), we can get:
\begin{equation*}
\begin{aligned}
   \mu_{\bm{d}_j} &= E[Y_i^{\bm{d}_j}] = E[R^{d_{j1}}Y^{s(d_{j1},0)}+(1-R^{d_{j1}})\sum_g I(G_i=g) Y^{s(d_{j1},g,d_{j2})}]  \\
   &= \pi_{d_{j1}} \mu_{s(d_{j1},0)} + 
   (1-\pi_{d_{j1}}) \sum_g \pr(G_i=g|A_{1i}=d_{j1},R_i=0)E[Y_i|A_{1i}=d_{j1},R_i=0,G_i=g,A_{2i}=d_{j2}] \\
   &= \pi_{d_{j1}} \mu_{s(d_{j1},0)} + 
   (1-\pi_{d_{j1}}) \sum_g \pr(G_i=g|A_{1i}=d_{j1},R_i=0)\mu_{s(d_{j1},g,d_{j2})},
\end{aligned}
\end{equation*}   
where $\mu_s(d_{j1},g,d_{j2})=E[Y_i|A_{1i}=d_{j1},R_i=0,G_i=g,A_{2i}=d_{j2}]$.

The IPW estimator for the value of DTR $\bm{d}_j$ in SMART-EXAM is 

\begin{equation*}
    \begin{aligned}
       \hat{\mu}_{\bm{d}_j} &= \frac{\sum_{i=1}^N W_i^{\bm{d}_j} Y_i}{\sum_{i=1}^N W_i^{\bm{d}_j}},
    \end{aligned}
\end{equation*}

where $W_i^{\bm{d}_j}= \frac{I(A_{1i}=d_{j1})}{p_{1,d_{j1}}} \left\{ R_i+ \frac{(1-R_i)I(A_{2i}=d_{j2})}{\sum_g I(G_i=g) p_{2,d_{j2}|d_{j1}, g}} \right\}$, and $p_{2,d_{j2}|d_{j1},g}$ is the randomization probability for treatment $d_{j2}$ when the initial treatment is $d_{j1}$ and in the g-th group. 

By the weak law of large numbers,

\begin{equation*}
    \begin{aligned}
    &\frac{\sum_{i=1}^N W_i^{\bm{d}_j}}{N} \xrightarrow{p} E[W_i^{\bm{d}_j}]= E\left[\frac{I(A_{1i}=d_{j1})}{p_{1,d_{j1}}} \left\{ R_i+ \frac{(1-R_i)I(A_{2i}=d_{j2})}{\sum_g I(G_i=g) p_{2,d_{j2}|d_{j1},g}} \right\} \right] \\
    &= E\left[E\left[\frac{I(A_{1i}=d_{j1})}{p_{1,d_{j1}}} \left\{ R_i+ \frac{(1-R_i)I(A_{2i}=d_{j2})}{\sum_g \{I(G_i=g) p_{2,d_{j2}|d_{j1},g}\}} \right\}\right]\right] \\
    &= \pr(A_{1i}=d_{j1},R_i=1)\frac{1}{p_{1,d_{j1}}}+ \sum_g \pr(A_{1i}=d_{j1},R_i=0,G_i=g,A_{2i}=d_{j2}) E\left[\frac{1}{p_{1,d_{j1}}
     p_{2,d_{j2}|d_{j1},g}}\right]\\
    &=\pi_{d_{j1}} + \sum_g p_{1,d_{j1}} (1-\pi_{d_{j1}}) \pr(G_i=g|A_{1i}=d_{j1},R_i=0) p_{2,d_{j2}|d_{j1},g}\frac{1}{p_{1,d_{j1}}p_{2,d_{j2}|d_{j1},g}} \\
    &= \pi_{d_{j1}}+(1-\pi_{d_{j1}})\\
    &= 1.
    \end{aligned}
\end{equation*}       

According to the weak law of large numbers and the continuous mapping theorem,

\begin{equation*}
    \begin{aligned}
     \hat{\mu}_{\bm{d}_j} &= \frac{\sum_{i=1}^N W_i^{\bm{d}_j} Y_i}{\sum_{i=1}^N W_i^{\bm{d}_j}}= \frac{(\sum_{i=1}^N W_i^{\bm{d}_j} Y_i)/N}{(\sum_{i=1}^N W_i^{\bm{d}_j})/N} \xrightarrow{p} \frac{E[W^{\bm{d}_j} Y]}{E[W^{\bm{d}_j}]} \\
     &= E[W^{\bm{d}_j} Y] = E\left[\frac{I(A_{1i}=d_{j1})}{p_{1,d_{j1}}} \left\{ R_i+ \frac{(1-R_i)I(A_{2i}=d_{j2})}{\sum_g I(G_i=g) p_{2,d_{j2}|d_{j1},g}} \right\} Y_i\right] \\
     &= E \left[ E\left[\frac{I(A_{1i}=d_{j1})}{p_{1,d_{j1}}} \left\{ R_i+ \frac{(1-R_i)I(A_{2i}=d_{j2})}{\sum_g I(G_i=g) p_{2,d_{j2}|d_{j1},g}} \right\} Y_i|A_{1i},R_i,G_i,A_{2i}\right] \right] \\
     &= \pr(A_{1i}=d_{j1},R_i=1)E\left[\frac{Y_i}{p_{1,d_{j1}}}|A_{1i}=d_{j1},R_i=1\right] \\
     &+ \sum_g \pr(A_{1i}=d_{j1},R_i=0,G_i=g,A_{2i}=d_{j2})E\left[\frac{Y_i}{p_{1,d_{j1}} p_{2,d_{j2}|d_{j1},g}}|A_{1i}=d_{j1},R_i=0,G_i=g,A_{2i}=d_{j2}\right] \\
     &= \pi_{d_{j1}}\mu_{s(d_{j1},0)} + \sum_g (1-\pi_{d_{j1}})\pr(G_i=g|A_{1i}=d_{j1},R_i=0)E[Y_i|A_{1i}=d_{j1},R_i=0,G_i=g,A_{2i}=d_{j2}] \\
     &= \mu_{\bm{d}_j}.
    \end{aligned}
\end{equation*}
\end{proof}

\begin{property}
\label{property_3}
Under SEA, CA, and PA, the large-sample distribution of $\hat{\mu}_{\bm{d}_j}$ for DTR $\bm{d}_j=(d_{j1},d_{j2})$ is 
\begin{equation}
\sqrt{N}(\hat{\mu}_{\bm{d}_j} - \mu_{\bm{d}_j}) \xrightarrow{d} N(0, \sigma^2_{\bm{d}_j}),
\end{equation}
where
\begin{equation}
\begin{aligned}
 \sigma^2_{\bm{d}_j}&=\frac{\pi_{1, d_{j1}}\{ \sigma^2_{s(d_{j1}, 0)}
 +(\mu_{\bm{d}_j}-\mu_{s(d_{j1},0)})^2 \}}{p_{d_{j1}}} \\
 & + \sum_{g} \left[ \frac{(1-\pi_{d_{j1}})\pr(G=g|A_1=d_{j1},R=0)\{ \sigma^2_{s(d_{j1},g,d_{j1})}+(\mu_{\bm{d}_j}-\mu_{s(d_{j1},g,d_{j2})})^2 \}}{p_{1,d_{j1}} (p_{2,d_{j2}|a_1,g})}\right],
\end{aligned}
\end{equation}
with $\mu_{s(d_{j1},g,d_{j2})}=E[Y|A_{1}=d_{j1}, R=0,G=g, A_{2}=d_{j2}]$, $\sigma^2_{s(d_{j1},g,d_{j2})}=\Var(Y|A_{1}=d_{j1}, R=0, G=g, A_{2}=d_{j2})$, $\mu_{s(d_{j1},0)}=E[Y|A_{1}=d_{j1}, R=1]$, $\sigma^2_{s(d_{j1},0)}=\Var(Y|A_{1}=d_{j1}, R=1)$.
\end{property}

\begin{proof}
    First, we derive the influence function of $\hat{\mu}_{\bm{d}_j}$ as follows:

We know that $\hat{\mu}_{\bm{d}_j}$ satisfies $g(\hat{\mu}_{\bm{d}_j})=\frac{1}{N} \sum_{i=1}^N W_i^{\bm{d}_j} (Y_i-\hat{\mu}_{\bm{d}_j})=0$. Expanding it with respect to $\mu_{\bm{d}_j}$, we can get
     $$\frac{1}{N} \sum_{i=1}^N W_i^{\bm{d}_j} (Y_i-\mu_{\bm{d}_j})-(\hat{\mu}_{\bm{d}_j}-\mu_{\bm{d}_j}) \frac{1}{N} \sum_{i=1}^N W_i^{\bm{d}_j} = 0$$
     $$\rightarrow \frac{1}{N} \sum_{i=1}^N W_i^{\bm{d}_j} (Y_i-\mu_{\bm{d}_j})-(\hat{\mu}_{\bm{d}_j}-\mu_{\bm{d}_j}) \frac{1}{N} (\sum_{i=1}^N W_i^{\bm{d}_j}-1)-(\hat{\mu}_{\bm{d}_j}-\mu_{\bm{d}_j})=0 $$
     $$\rightarrow \hat{\mu}_{\bm{d}_j}-\mu_{\bm{d}_j}=\frac{1}{N} \sum_{i=1}^N W_i^{\bm{d}_j} (Y_i-\mu_{\bm{d}_j})-(\hat{\mu}_{\bm{d}_j}-\mu_{\bm{d}_j}) \frac{1}{N} \sum_{i=1}^N (W_i^{\bm{d}_j}-1)$$
     $$\rightarrow \sqrt{N} (\hat{\mu}_{\bm{d}_j}-\mu_{\bm{d}_j})=N^{-1/2} \sum_{i=1}^N W_i^{\bm{d}_j} (Y_i-\mu_{\bm{d}_j})-(\hat{\mu}_{\bm{d}_j}-\mu_{\bm{d}_j}) N^{-1/2} \sum_{i=1}^n (W_i^{\bm{d}_j}-1)$$
     $$\rightarrow \sqrt{N} (\hat{\mu}_{\bm{d}_j}-\mu_{\bm{d}_j})=N^{-1/2} \sum_{i=1}^N \psi_i^{\bm{d}_j}-(\hat{\mu}_{\bm{d}_j}-\mu_{\bm{d}_j}) N^{-1/2} \sum_{i=1}^N (W_i^{\bm{d}_j}-1).$$

According to the properties of M-estimators and the central limit theorem, we can get 
$$\sqrt{N} (\hat{\mu}_{\bm{d}_j}-\mu_{\bm{d}_j}) \xrightarrow{d} N(0, \sigma^2_{\bm{d}_j}), $$
where 
$$\sigma^2_{\bm{d}_j}=Var(\psi_i^{\bm{d}_j})=E[(\psi_i^{\bm{d}_j})^2]=E[(W_i^{\bm{d}_j})^2 (Y_i-\mu_{\bm{d}_j})^2].$$
The variance of the DTR mean estimator is derived as follows:
\begin{equation*}
    \begin{aligned}
     \sigma^2_{\bm{d}_j} &= E[(W_i^{\bm{d}_j})^2(Y_i-\mu_{\bm{d}_j})^2]\\
     &= E\left[E\left[\frac{I(A_{1i}=d_{j1})}{p^2_{1,d_{j1}}} \left\{ R_i+ \frac{(1-R_i)I(A_{2i}=a_2)}{\sum_g I(G_i=g) p_{2,a_2|d_{j1},g}} \right\}^2 (Y_i-\mu_{\bm{d}_j})^2|A_{1i}, R_i, G_i, A_{2i}\right]\right] \\
     &= \pr(A_{1i}=d_{j1}, R_i=1) E[\frac{(Y_i-\mu_{\bm{d}_j})^2}{p^2_{1,d_{j1}}}|A_{1i}=d_{j1}, R_i=1] \\
     &+ \sum_g  \left\{ \pr(A_{1i}=d_{j1}, R_i=0, G_i=g, A_{2i}=d_{j2}) \times E\left[\frac{(Y_i-\mu_{\bm{d}_j})^2}{p^2_{1,d_{j1}}p^2_{2,d_{j1}|d_{j1},g}}|A_{1i}=d_{j1}, R_i=0, G_i=g, A_{2i}=d_{j2}\right] \right\} \\
     &= \frac{\pi_{d_{j1}}}{p_{1,d_{j1}}} (\mu^2_{s(d_{j1},0)}+ \sigma^2_{s(d_{j1},0)}+\mu_{\bm{d}_j}^2-2\mu_{\bm{d}_j} \mu_{s(d_{j1},0)} ) \\
     &+ \sum_g \left\{ \frac{(1-\pi_{d_{j1}})\pr(G_i=g|A_{1i}=d_{j1},R_i=0)}{ p_{1,d_{j1}} (p_{2,d_{j2}|d_{j1},g})}  (\mu^2_{s(d_{j1},g,d_{j2})}+ \sigma^2_{s(d_{j1},g,d_{j2})}  +\mu_{\bm{d}_j}^2 -2\mu_{\bm{d}_j}\mu_{s(d_{j1},g,d_{j2})}) \right\} \\
     &= \frac{\pi_{d_{j1}}}{p_{1,d_{j1}}} \left\{ (\mu_{s(d_{j1},0)}-\mu_{\bm{d}_j})^2 + \sigma^2_{s(d_{j1},0)} \right\} \\
     &+ \sum_g \left[ \frac{(1-\pi_{d_{j1}})\pr(G_i=g|A_{1i}=d_{j1},R_i=0)}{ p_{1,d_{j1}} (p_{2,d_{j2}|d_{j1},g})}   \left\{  (\mu_{s(d_{j1},g,d_{j2})}-\mu_{\bm{d}_j})^2 + \sigma^2_{s(d_{j1},g,d_{j2})} \right\} \right], 
    \end{aligned}
\end{equation*}
where $\mu_{s(d_{j1},g,d_{j2})}=E[Y_i|A_{1i}=d_{j1}, R_i=0,G_i=g, A_{2i}=d_{j2}]$, $\sigma^2_{s(d_{j1},g,d_{j2})}=\Var(Y_i|A_{1i}=d_{j1}, R_i=0,G_i=g, A_{2i}=d_{j2})$, $\mu_{s(d_{j1},0)}=E[Y_i|A_{1i}=d_{j1}, R_i=1]$, $\sigma^2_{s(d_{j1},0)}=\Var(Y_i|A_{1i}=d_{j1}, R_i=1)$.

So,
\begin{equation*}
    \begin{aligned}
     \Var(\mu_{\bm{d}_j}) &=  \sigma^2_{\bm{d}_j}/N \\
     &= 1/N(\frac{\pi_{d_{j1}}}{p_{1,d_{j1}}} \{\sigma^2_{s(d_{j1},0)}+(\mu_{\bm{d}_j}-\mu_{s(d_{j1},0)})^2\} \\
     &+ \sum_g \left[\frac{(1-\pi_{d_{j1}})\pr (G=g|A_{1}=d_{j1},R=0)}{ p_{1,d_{j1}} (p_{2,d_{j1}|a_1,g})} \}  \left\{\sigma^2_{s(d_{j1},g, d_{j2})} + (\mu_{\bm{d}_j}-\mu_{s(d_{j1},g,d_{j2})})^2 \right\} \right].
    \end{aligned}
\end{equation*}

The covariance of the outcome means for the DTRs that share with the same initial treatment. i.e., $\mu_{\bm{d}_{j}}=\mu_{(d_{j1},d_{j2})}$ and $\mu_{\bm{d}'_j}=\mu_{(d_{j1},d'_{j2})}$, can be derived by
\begin{equation*}
    \begin{aligned}
     \sigma_{\bm{d}_j,\bm{d}'_j} &= \cov(\psi_{\bm{d}_j},\psi_{\bm{d}'_j})=\cov ( W_i^{\bm{d}_j}(Y_i-\mu_{\bm{d}_j}),W_i^{\bm{d}'_j} (Y_i-\mu_{\bm{d}'_j}) ) \\
     &=E\left[W_i^{\bm{d}_j}W_i^{\bm{d}'_j}(Y_i-\mu_{\bm{d}_j})(Y_i-\mu_{\bm{d}'_j})\right]\\
     &= P(A_{1i}=d_{j1}, R_i=1) E\left[\frac{(Y_i-\mu_{\bm{d}_j})(Y_i-\mu_{\bm{d}'_j})}{p^2_{1,d_{j1}}}|A_{1i}=d_{j1}, R_i=1\right]  \\
     &= \frac{\pi_{d_{j1}}}{p_{1,d_{j1}}} (E[Y_i^2|A_{1i}=d_{j1}, R_i=1]+\mu_{\bm{d}_j}\mu_{\bm{d}'_j} -\mu_{\bm{d}_j}E[Y_i|A_{1i}=d_{j1}, R_i=1]-\mu_{\bm{d}'_j}E[Y_i|A_{1i}=d_{j1}, R_i=1] ) \\
     &= \frac{\pi_{d_{j1}}}{p_{1,d_{j1}}} \left\{ \sigma^2_{s(d_{j1},0)}+ (\mu_{s(d_{j1},0)}- \mu_{\bm{d}_j})(\mu_{s(d_{j1},0)}-\mu_{\bm{d}'_j}) \right\}.
    \end{aligned}
\end{equation*}
So,
\begin{equation*}
    \begin{aligned}
     \cov(\mu_{\bm{d}_j},\mu_{\bm{d}'_j}) &=  \frac{1}{N} \sigma_{\bm{d}_j,\bm{d}'_j} = \frac{1}{N} \times \frac{\pi_{d_{j1}}}{p_{1,d_{j1}}} \left\{ \sigma^2_{s(d_{j1},0)}+ (\mu_{s(d_{j1},0)}- \mu_{\bm{d}_j})(\mu_{s(d_{j1},0)}-\mu_{\bm{d}'_j}) \right\}.
    \end{aligned}
\end{equation*}
\end{proof}

\newpage

\section*{Web Appendix D: Tables and figures}
\renewcommand{\figurename}{Web Figure}
\renewcommand{\tablename}{Web Table}

\begin{figure}[H]
\centering
\caption{Simulation results for the setting with a pilot SMART of size $N^p=100$. I, II, III, and IV correspond to the sample size $N=200, 300, 400, 500$, respectively. The left (right) panel of each subfigure corresponds to the negative (positive) association between preferences and treatment effects. $E[\mu_{\hat{\bm{d}}^{\ast}}]$ denotes the Monte Carlo Mean of the true value of the estimated optimal DTR; $E[\bar{u}]$ denotes the Monte Carlo Mean of the probability of being assigned to the preferred treatment, and $E[\bar{Y}]$ denotes the Monte Carlo Mean of the mean outcome.}
\label{simu_plot_1}
\includegraphics[width=\linewidth]{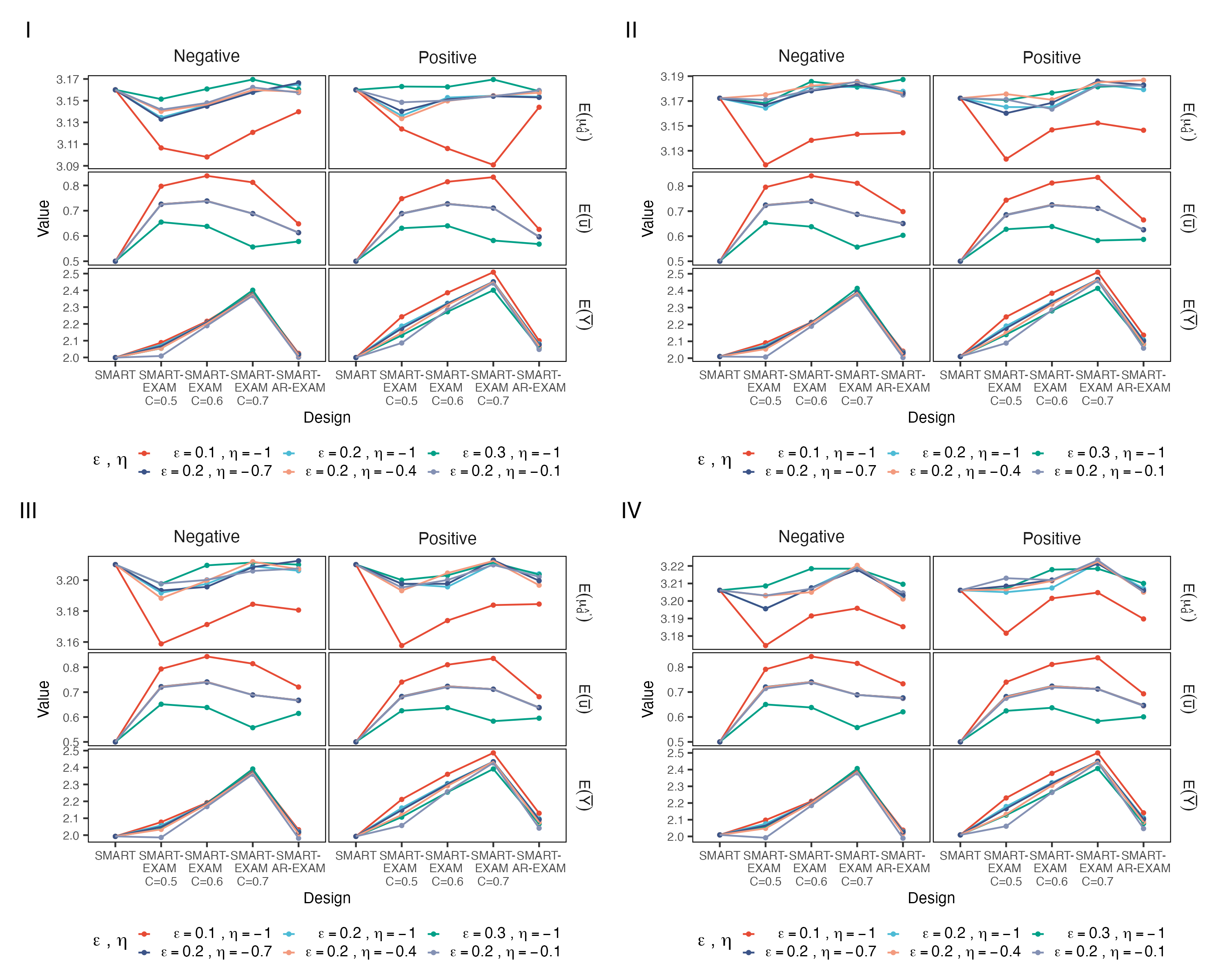}
\end{figure}

\begin{figure}[H]
\centering
\caption{Simulation results for the setting with a pilot SMART of size $N^p=300$. I, II, III, and IV correspond to the sample size $N=200, 300, 400, 500$, respectively. The left (right) panel of each subfigure corresponds to the negative (positive) association between preferences and treatment effects. $E[\mu_{\hat{\bm{d}}^{\ast}}]$ denotes the Monte Carlo Mean of the true value of the estimated optimal DTR; $E[\bar{u}]$ denotes the Monte Carlo Mean of the probability of being assigned to the preferred treatment, and $E[\bar{Y}]$ denotes the Monte Carlo Mean of the mean outcome.}
\label{simu_plot_2}
\includegraphics[width=\linewidth]{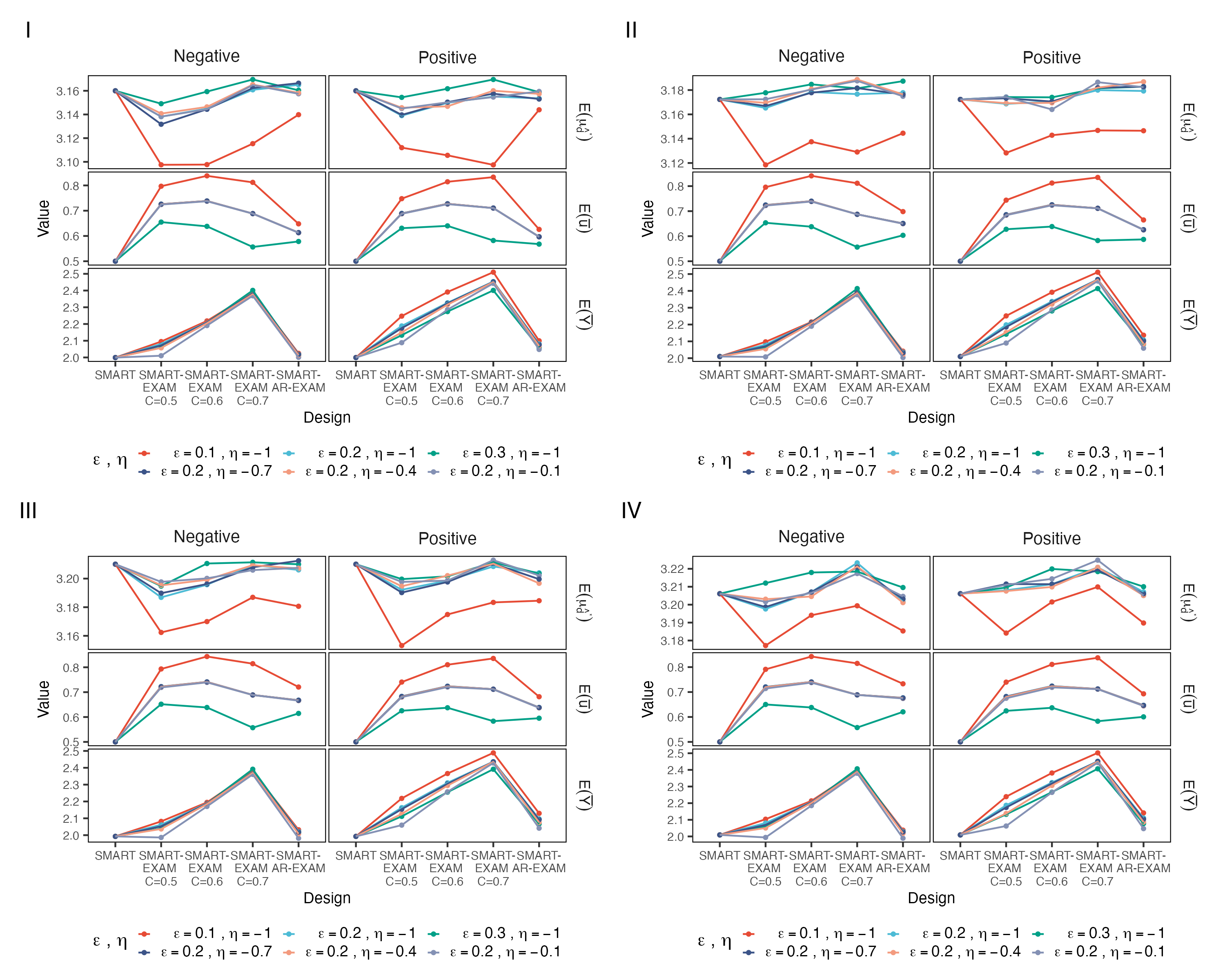}
\end{figure}

\begin{table}
 \centering
 \caption{ The operating characteristics of each SMART design when the total sample size is $N=200$, the sample size for the pilot SMART is $N^p=200$, $\epsilon=0.1$, and $\eta=-1$. $\mu_{\bm{d}_j}$ denotes the true value of DTR $\bm{d}_j$;  $E[\hat{\mu}^{IPW}_{\bm{d}_j}]$ and $E[\hat{\mu}^{AIPW}_{\bm{d}_j}]$ denote the Monte Carlo mean of the estimated DTR mean by IPW and AIPW, respectively;  $\se (\hat{\mu}^{IPW}_{\bm{d}_j})$ and $\se (\hat{\mu}^{AIPW}_{\bm{d}_j})$ denote the empirical standard error of the estimated DTR mean by IPW and AIPW, respectively; $\bar{N}_{\bm{d}_j}$ is the Monte Carlo mean of the number of participants in DTR $\bm{d}_j$; $\se (N_{\bm{d}_j})$ denotes the empirical standard error of the number of participants in DTR $\bm{d}_j$; $\Pr(\bm{d}_j=\hat{\bm{d}}^{\ast IPW})$ and $\Pr(\bm{d}_j=\hat{\bm{d}}^{\ast AIPW})$ correspond to the probability of being selected as the optimal DTR for each DTR $\bm{d}_j$ by IPW and AIPW, respectively.}
 \label{table_1}
 \resizebox{0.7\columnwidth}{!}{%
\begin{tabular}{lccccccccc}
  \hline
   \multicolumn{2}{c}{DTR} & \multicolumn{8}{c}{SMART, Negative}  \\
  \cmidrule(lr){1-2}\cmidrule(lr){3-10}
DTR & $\mu_{\bm{d}_j}$ & 
$E[\hat{\mu}^{IPW}_{\bm{d}_j}]$ & $\se (\hat{\mu}^{IPW}_{\bm{d}_j})$ & $E[\hat{\mu}^{AIPW}_{\bm{d}_j}]$ & $\se (\hat{\mu}^{AIPW}_{\bm{d}_j})$ & $\bar{N}_{\bm{d}_j}$ & $\se (N_{\bm{d}_j})$ & $\Pr(\bm{d}_j=\hat{\bm{d}}^{\ast IPW})$ & $\Pr(\bm{d}_j=\hat{\bm{d}}^{\ast AIPW})$ \\ 
  \hline
(-1,-1) & 2.234 & 2.22 & 0.36 & 2.222 & 0.362 & 74.936 & 6.654 & 0.008 & 0.008 \\ 
  (-1,1) & 2.758 & 2.756 & 0.375 & 2.759 & 0.375 & 75.056 & 6.848 & 0.166 & 0.166 \\ 
  (1,-1) & 1.76 & 1.777 & 0.461 & 1.766 & 0.429 & 75.244 & 6.739 & 0.002 & 0 \\ 
  (1,1) & 3.251 & 3.28 & 0.382 & 3.274 & 0.382 & 75.162 & 6.816 & 0.824 & 0.826 \\ 
   \hline
   \multicolumn{2}{c}{DTR} & \multicolumn{8}{c}{SMART, Positive}  \\
  \cmidrule(lr){1-2}\cmidrule(lr){3-10}
DTR & $\mu_{\bm{d}_j}$ & 
$E[\hat{\mu}^{IPW}_{\bm{d}_j}]$ & $\se (\hat{\mu}^{IPW}_{\bm{d}_j})$ & $E[\hat{\mu}^{AIPW}_{\bm{d}_j}]$ & $\se (\hat{\mu}^{AIPW}_{\bm{d}_j})$ & $\bar{N}_{\bm{d}_j}$ & $\se (N_{\bm{d}_j})$ & $\Pr(\bm{d}_j=\hat{\bm{d}}^{\ast IPW})$ & $\Pr(\bm{d}_j=\hat{\bm{d}}^{\ast AIPW})$ \\ 
  \hline
  (-1,-1) & 2.234 & 2.22 & 0.36 & 2.222 & 0.362 & 74.936 & 6.654 & 0.008 & 0.008 \\ 
  (-1,1) & 2.758 & 2.756 & 0.375 & 2.759 & 0.375 & 75.056 & 6.848 & 0.166 & 0.166 \\ 
  (1,-1) & 1.76 & 1.777 & 0.461 & 1.766 & 0.429 & 75.244 & 6.739 & 0.002 & 0 \\ 
  (1,1) & 3.251 & 3.28 & 0.382 & 3.274 & 0.382 & 75.162 & 6.816 & 0.824 & 0.826 \\ 
  \hline
  \multicolumn{2}{c}{DTR} &
  \multicolumn{8}{c}{SMART-EXAM $(C=0.5)$, Negative} \\
  \cmidrule(lr){1-2}\cmidrule(lr){3-10}
DTR & $\mu_{\bm{d}_j}$ & 
$E[\hat{\mu}^{IPW}_{\bm{d}_j}]$ & $\se (\hat{\mu}^{IPW}_{\bm{d}_j})$ & $E[\hat{\mu}^{AIPW}_{\bm{d}_j}]$ & $\se (\hat{\mu}^{AIPW}_{\bm{d}_j})$ & $\bar{N}_{\bm{d}_j}$ & $\se (N_{\bm{d}_j})$ & $\Pr(\bm{d}_j=\hat{\bm{d}}^{\ast IPW})$ & $\Pr(\bm{d}_j=\hat{\bm{d}}^{\ast AIPW})$ \\ 
  \hline
(-1,-1) & 2.234 & 2.228 & 0.478 & 2.219 & 0.487 & 74.508 & 6.305 & 0.03 & 0.036 \\ 
  (-1,1) & 2.758 & 2.73 & 0.506 & 2.747 & 0.502 & 75.484 & 6.246 & 0.214 & 0.218 \\ 
  (1,-1) & 1.76 & 1.8 & 0.657 & 1.744 & 0.578 & 74.734 & 6.256 & 0.014 & 0.004 \\ 
  (1,1) & 3.251 & 3.268 & 0.517 & 3.257 & 0.504 & 75.672 & 6.283 & 0.742 & 0.742 \\
   \hline
\multicolumn{2}{c}{DTR} &
  \multicolumn{8}{c}{SMART-EXAM $(C=0.5)$, Positive} \\
  \cmidrule(lr){1-2}\cmidrule(lr){3-10}
DTR & $\mu_{\bm{d}_j}$ & 
$E[\hat{\mu}^{IPW}_{\bm{d}_j}]$ & $\se (\hat{\mu}^{IPW}_{\bm{d}_j})$ & $E[\hat{\mu}^{AIPW}_{\bm{d}_j}]$ & $\se (\hat{\mu}^{AIPW}_{\bm{d}_j})$ & $\bar{N}_{\bm{d}_j}$ & $\se (N_{\bm{d}_j})$ & $\Pr(\bm{d}_j=\hat{\bm{d}}^{\ast IPW})$ & $\Pr(\bm{d}_j=\hat{\bm{d}}^{\ast AIPW})$ \\ 
  \hline
(-1,-1) & 2.234 & 2.219 & 0.483 & 2.21 & 0.488 & 74.544 & 6.247 & 0.022 & 0.022 \\ 
  (-1,1) & 2.758 & 2.75 & 0.506 & 2.756 & 0.5 & 75.448 & 6.275 & 0.216 & 0.19 \\ 
  (1,-1) & 1.76 & 1.795 & 0.613 & 1.726 & 0.532 & 74.494 & 6.425 & 0.004 & 0 \\ 
  (1,1) & 3.251 & 3.285 & 0.484 & 3.275 & 0.468 & 75.912 & 6.576 & 0.758 & 0.788 \\
   \hline
   
\multicolumn{2}{c}{DTR} &
  \multicolumn{8}{c}{SMART-EXAM $(C=0.6)$, Negative} \\
  \cmidrule(lr){1-2}\cmidrule(lr){3-10}
DTR & $\mu_{\bm{d}_j}$ & 
$E[\hat{\mu}^{IPW}_{\bm{d}_j}]$ & $\se (\hat{\mu}^{IPW}_{\bm{d}_j})$ & $E[\hat{\mu}^{AIPW}_{\bm{d}_j}]$ & $\se (\hat{\mu}^{AIPW}_{\bm{d}_j})$ & $\bar{N}_{\bm{d}_j}$ & $\se (N_{\bm{d}_j})$ & $\Pr(\bm{d}_j=\hat{\bm{d}}^{\ast IPW})$ & $\Pr(\bm{d}_j=\hat{\bm{d}}^{\ast AIPW})$ \\ 
  \hline
(-1,-1) & 2.234 & 2.212 & 0.542 & 2.215 & 0.553 & 69.52 & 6.161 & 0.038 & 0.04 \\ 
  (-1,1) & 2.758 & 2.754 & 0.448 & 2.757 & 0.425 & 80.472 & 6.262 & 0.182 & 0.17 \\ 
  (1,-1) & 1.76 & 1.823 & 0.727 & 1.753 & 0.622 & 70.074 & 6.093 & 0.018 & 0.012 \\ 
  (1,1) & 3.251 & 3.252 & 0.443 & 3.253 & 0.437 & 80.332 & 6.532 & 0.762 & 0.778 \\ 
   \hline
\multicolumn{2}{c}{DTR} &
  \multicolumn{8}{c}{SMART-EXAM $(C=0.6)$, Positive} \\
  \cmidrule(lr){1-2}\cmidrule(lr){3-10}
DTR & $\mu_{\bm{d}_j}$ & 
$E[\hat{\mu}^{IPW}_{\bm{d}_j}]$ & $\se (\hat{\mu}^{IPW}_{\bm{d}_j})$ & $E[\hat{\mu}^{AIPW}_{\bm{d}_j}]$ & $\se (\hat{\mu}^{AIPW}_{\bm{d}_j})$ & $\bar{N}_{\bm{d}_j}$ & $\se (N_{\bm{d}_j})$ & $\Pr(\bm{d}_j=\hat{\bm{d}}^{\ast IPW})$ & $\Pr(\bm{d}_j=\hat{\bm{d}}^{\ast AIPW})$ \\ 
  \hline
(-1,-1) & 2.234 & 2.22 & 0.563 & 2.214 & 0.573 & 69.596 & 6.103 & 0.042 & 0.046 \\ 
  (-1,1) & 2.758 & 2.754 & 0.447 & 2.754 & 0.43 & 80.396 & 6.263 & 0.184 & 0.184 \\ 
  (1,-1) & 1.76 & 1.833 & 0.698 & 1.743 & 0.609 & 69.436 & 6.202 & 0.012 & 0.002 \\ 
  (1,1) & 3.251 & 3.254 & 0.424 & 3.249 & 0.417 & 80.97 & 6.447 & 0.762 & 0.768 \\
   \hline
   
 \multicolumn{2}{c}{DTR} &
  \multicolumn{8}{c}{SMART-EXAM $(C=0.7)$, Negative} \\
  \cmidrule(lr){1-2}\cmidrule(lr){3-10}
DTR & $\mu_{\bm{d}_j}$ & 
$E[\hat{\mu}^{IPW}_{\bm{d}_j}]$ & $\se (\hat{\mu}^{IPW}_{\bm{d}_j})$ & $E[\hat{\mu}^{AIPW}_{\bm{d}_j}]$ & $\se (\hat{\mu}^{AIPW}_{\bm{d}_j})$ & $\bar{N}_{\bm{d}_j}$ & $\se (N_{\bm{d}_j})$ & $\Pr(\bm{d}_j=\hat{\bm{d}}^{\ast IPW})$ & $\Pr(\bm{d}_j=\hat{\bm{d}}^{\ast AIPW})$ \\   
  \hline
(-1,-1) & 2.234 & 2.203 & 0.592 & 2.212 & 0.605 & 65.15 & 6.062 & 0.038 & 0.044 \\ 
  (-1,1) & 2.758 & 2.755 & 0.376 & 2.757 & 0.365 & 84.842 & 6.482 & 0.138 & 0.146 \\ 
  (1,-1) & 1.76 & 1.824 & 0.772 & 1.745 & 0.662 & 65.634 & 6.232 & 0.02 & 0.01 \\ 
  (1,1) & 3.251 & 3.263 & 0.377 & 3.262 & 0.367 & 84.772 & 6.679 & 0.804 & 0.8 \\ 
   \hline
   \multicolumn{2}{c}{DTR} &
  \multicolumn{8}{c}{SMART-EXAM $(C=0.7)$, Positive} \\
  \cmidrule(lr){1-2}\cmidrule(lr){3-10}
DTR & $\mu_{\bm{d}_j}$ & 
$E[\hat{\mu}^{IPW}_{\bm{d}_j}]$ & $\se (\hat{\mu}^{IPW}_{\bm{d}_j})$ & $E[\hat{\mu}^{AIPW}_{\bm{d}_j}]$ & $\se (\hat{\mu}^{AIPW}_{\bm{d}_j})$ & $\bar{N}_{\bm{d}_j}$ & $\se (N_{\bm{d}_j})$ & $\Pr(\bm{d}_j=\hat{\bm{d}}^{\ast IPW})$ & $\Pr(\bm{d}_j=\hat{\bm{d}}^{\ast AIPW})$ \\   
  \hline
(-1,-1) & 2.234 & 2.215 & 0.586 & 2.213 & 0.599 & 65.376 & 6.094 & 0.046 & 0.054 \\ 
  (-1,1) & 2.758 & 2.756 & 0.38 & 2.754 & 0.369 & 84.616 & 6.451 & 0.16 & 0.164 \\ 
  (1,-1) & 1.76 & 1.868 & 0.78 & 1.755 & 0.678 & 64.758 & 6.075 & 0.03 & 0.014 \\ 
  (1,1) & 3.251 & 3.249 & 0.371 & 3.25 & 0.364 & 85.648 & 6.775 & 0.764 & 0.768 \\
   \hline
   \multicolumn{2}{c}{DTR} &
  \multicolumn{8}{c}{SMART-AR-EXAM $(C=0.5)$, Negative}  \\
  \cmidrule(lr){1-2}\cmidrule(lr){3-10}
DTR & $\mu_{\bm{d}_j}$ & 
$E[\hat{\mu}^{IPW}_{\bm{d}_j}]$ & $\se (\hat{\mu}^{IPW}_{\bm{d}_j})$ & $E[\hat{\mu}^{AIPW}_{\bm{d}_j}]$ & $\se (\hat{\mu}^{AIPW}_{\bm{d}_j})$ & $\bar{N}_{\bm{d}_j}$ & $\se (N_{\bm{d}_j})$ & $\Pr(\bm{d}_j=\hat{\bm{d}}^{\ast IPW})$ & $\Pr(\bm{d}_j=\hat{\bm{d}}^{\ast AIPW})$ \\  
  \hline
(-1,-1) & 2.234 & 2.251 & 0.421 & 2.254 & 0.426 & 74.538 & 6.573 & 0.02 & 0.022 \\ 
  (-1,1) & 2.758 & 2.712 & 0.419 & 2.724 & 0.421 & 75.454 & 6.393 & 0.16 & 0.182 \\ 
  (1,-1) & 1.76 & 1.835 & 0.534 & 1.804 & 0.492 & 74.894 & 6.412 & 0.006 & 0.002 \\ 
  (1,1) & 3.251 & 3.256 & 0.444 & 3.241 & 0.447 & 75.512 & 6.576 & 0.814 & 0.794 \\
   \hline
     \multicolumn{2}{c}{DTR} &
  \multicolumn{8}{c}{SMART-AR-EXAM $(C=0.5)$, Positive}  \\
  \cmidrule(lr){1-2}\cmidrule(lr){3-10}
DTR & $\mu_{\bm{d}_j}$ & 
$E[\hat{\mu}^{IPW}_{\bm{d}_j}]$ & $\se (\hat{\mu}^{IPW}_{\bm{d}_j})$ & $E[\hat{\mu}^{AIPW}_{\bm{d}_j}]$ & $\se (\hat{\mu}^{AIPW}_{\bm{d}_j})$ & $\bar{N}_{\bm{d}_j}$ & $\se (N_{\bm{d}_j})$ & $\Pr(\bm{d}_j=\hat{\bm{d}}^{\ast IPW})$ & $\Pr(\bm{d}_j=\hat{\bm{d}}^{\ast AIPW})$ \\  
  \hline
(-1,-1) & 2.234 & 2.251 & 0.425 & 2.249 & 0.423 & 74.65 & 6.601 & 0.016 & 0.016 \\ 
  (-1,1) & 2.758 & 2.718 & 0.437 & 2.731 & 0.437 & 75.342 & 6.443 & 0.178 & 0.186 \\ 
  (1,-1) & 1.76 & 1.823 & 0.522 & 1.779 & 0.467 & 74.598 & 6.372 & 0.002 & 0 \\ 
  (1,1) & 3.251 & 3.256 & 0.433 & 3.249 & 0.432 & 75.808 & 6.757 & 0.804 & 0.798 \\ 
   \hline
\end{tabular}
}
\end{table}

\begin{table}
 \centering
 \caption{ The operating characteristics of each SMART design when the total sample size is $N=200$, the sample size for the pilot SMART is $N^p=200$, $\epsilon=0.2$, and $\eta=-1$. $\mu_{\bm{d}_j}$ denotes the true value of DTR $\bm{d}_j$;  $E[\hat{\mu}^{IPW}_{\bm{d}_j}]$ and $E[\hat{\mu}^{AIPW}_{\bm{d}_j}]$ denote the Monte Carlo mean of the estimated DTR mean by IPW and AIPW, respectively;  $\se (\hat{\mu}^{IPW}_{\bm{d}_j})$ and $\se (\hat{\mu}^{AIPW}_{\bm{d}_j})$ denote the empirical standard error of the estimated DTR mean by IPW and AIPW, respectively; $\bar{N}_{\bm{d}_j}$ is the Monte Carlo mean of the number of participants in DTR $\bm{d}_j$; $\se (N_{\bm{d}_j})$ denotes the empirical standard error of the number of participants in DTR $\bm{d}_j$; $\Pr(\bm{d}_j=\hat{\bm{d}}^{\ast IPW})$ and $\Pr(\bm{d}_j=\hat{\bm{d}}^{\ast AIPW})$ correspond to the probability of being selected as the optimal DTR for each DTR $\bm{d}_j$ by IPW and AIPW, respectively.}
 \label{table_2}
 \resizebox{0.7\columnwidth}{!}{%
\begin{tabular}{lccccccccc}
  \hline
   \multicolumn{2}{c}{DTR} & \multicolumn{8}{c}{SMART, Negative}  \\
  \cmidrule(lr){1-2}\cmidrule(lr){3-10}
DTR & $\mu_{\bm{d}_j}$ & 
$E[\hat{\mu}^{IPW}_{\bm{d}_j}]$ & $\se (\hat{\mu}^{IPW}_{\bm{d}_j})$ & $E[\hat{\mu}^{AIPW}_{\bm{d}_j}]$ & $\se (\hat{\mu}^{AIPW}_{\bm{d}_j})$ & $\bar{N}_{\bm{d}_j}$ & $\se (N_{\bm{d}_j})$ & $\Pr(\bm{d}_j=\hat{\bm{d}}^{\ast IPW})$ & $\Pr(\bm{d}_j=\hat{\bm{d}}^{\ast AIPW})$ \\ 
  \hline
(-1,-1) & 2.234 & 2.22 & 0.36 & 2.222 & 0.362 & 74.936 & 6.654 & 0.008 & 0.008 \\ 
  (-1,1) & 2.758 & 2.756 & 0.375 & 2.759 & 0.375 & 75.056 & 6.848 & 0.166 & 0.166 \\ 
  (1,-1) & 1.76 & 1.777 & 0.461 & 1.766 & 0.429 & 75.244 & 6.739 & 0.002 & 0 \\ 
  (1,1) & 3.251 & 3.28 & 0.382 & 3.274 & 0.382 & 75.162 & 6.816 & 0.824 & 0.826 \\ 
   \hline
   \multicolumn{2}{c}{DTR} & \multicolumn{8}{c}{SMART, Positive}  \\
  \cmidrule(lr){1-2}\cmidrule(lr){3-10}
DTR & $\mu_{\bm{d}_j}$ & 
$E[\hat{\mu}^{IPW}_{\bm{d}_j}]$ & $\se (\hat{\mu}^{IPW}_{\bm{d}_j})$ & $E[\hat{\mu}^{AIPW}_{\bm{d}_j}]$ & $\se (\hat{\mu}^{AIPW}_{\bm{d}_j})$ & $\bar{N}_{\bm{d}_j}$ & $\se (N_{\bm{d}_j})$ & $\Pr(\bm{d}_j=\hat{\bm{d}}^{\ast IPW})$ & $\Pr(\bm{d}_j=\hat{\bm{d}}^{\ast AIPW})$ \\ 
  \hline
  (-1,-1) & 2.234 & 2.22 & 0.36 & 2.222 & 0.362 & 74.936 & 6.654 & 0.008 & 0.008 \\ 
  (-1,1) & 2.758 & 2.756 & 0.375 & 2.759 & 0.375 & 75.056 & 6.848 & 0.166 & 0.166 \\ 
  (1,-1) & 1.76 & 1.777 & 0.461 & 1.766 & 0.429 & 75.244 & 6.739 & 0.002 & 0 \\ 
  (1,1) & 3.251 & 3.28 & 0.382 & 3.274 & 0.382 & 75.162 & 6.816 & 0.824 & 0.826 \\
  \hline
  \multicolumn{2}{c}{DTR} &
  \multicolumn{8}{c}{SMART-EXAM $(C=0.5)$, Negative} \\
  \cmidrule(lr){1-2}\cmidrule(lr){3-10}
DTR & $\mu_{\bm{d}_j}$ & 
$E[\hat{\mu}^{IPW}_{\bm{d}_j}]$ & $\se (\hat{\mu}^{IPW}_{\bm{d}_j})$ & $E[\hat{\mu}^{AIPW}_{\bm{d}_j}]$ & $\se (\hat{\mu}^{AIPW}_{\bm{d}_j})$ & $\bar{N}_{\bm{d}_j}$ & $\se (N_{\bm{d}_j})$ & $\Pr(\bm{d}_j=\hat{\bm{d}}^{\ast IPW})$ & $\Pr(\bm{d}_j=\hat{\bm{d}}^{\ast AIPW})$ \\ 
  \hline
(-1,-1) & 2.234 & 2.213 & 0.407 & 2.211 & 0.411 & 74.33 & 6.486 & 0.016 & 0.016 \\ 
  (-1,1) & 2.758 & 2.765 & 0.427 & 2.772 & 0.417 & 75.662 & 6.484 & 0.196 & 0.206 \\ 
  (1,-1) & 1.76 & 1.787 & 0.547 & 1.754 & 0.491 & 74.576 & 6.437 & 0.002 & 0.002 \\ 
  (1,1) & 3.251 & 3.26 & 0.458 & 3.257 & 0.446 & 75.83 & 6.565 & 0.786 & 0.776 \\
   \hline
\multicolumn{2}{c}{DTR} &
  \multicolumn{8}{c}{SMART-EXAM $(C=0.5)$, Positive} \\
  \cmidrule(lr){1-2}\cmidrule(lr){3-10}
DTR & $\mu_{\bm{d}_j}$ & 
$E[\hat{\mu}^{IPW}_{\bm{d}_j}]$ & $\se (\hat{\mu}^{IPW}_{\bm{d}_j})$ & $E[\hat{\mu}^{AIPW}_{\bm{d}_j}]$ & $\se (\hat{\mu}^{AIPW}_{\bm{d}_j})$ & $\bar{N}_{\bm{d}_j}$ & $\se (N_{\bm{d}_j})$ & $\Pr(\bm{d}_j=\hat{\bm{d}}^{\ast IPW})$ & $\Pr(\bm{d}_j=\hat{\bm{d}}^{\ast AIPW})$ \\ 
  \hline
(-1,-1) & 2.234 & 2.21 & 0.406 & 2.209 & 0.408 & 74.336 & 6.515 & 0.01 & 0.014 \\ 
  (-1,1) & 2.758 & 2.775 & 0.43 & 2.779 & 0.413 & 75.656 & 6.445 & 0.192 & 0.192 \\ 
  (1,-1) & 1.76 & 1.766 & 0.545 & 1.741 & 0.484 & 74.556 & 6.508 & 0.002 & 0 \\ 
  (1,1) & 3.251 & 3.276 & 0.425 & 3.274 & 0.417 & 75.85 & 6.744 & 0.796 & 0.794 \\ 
   \hline
   
\multicolumn{2}{c}{DTR} &
  \multicolumn{8}{c}{SMART-EXAM $(C=0.6)$, Negative} \\
  \cmidrule(lr){1-2}\cmidrule(lr){3-10}
DTR & $\mu_{\bm{d}_j}$ & 
$E[\hat{\mu}^{IPW}_{\bm{d}_j}]$ & $\se (\hat{\mu}^{IPW}_{\bm{d}_j})$ & $E[\hat{\mu}^{AIPW}_{\bm{d}_j}]$ & $\se (\hat{\mu}^{AIPW}_{\bm{d}_j})$ & $\bar{N}_{\bm{d}_j}$ & $\se (N_{\bm{d}_j})$ & $\Pr(\bm{d}_j=\hat{\bm{d}}^{\ast IPW})$ & $\Pr(\bm{d}_j=\hat{\bm{d}}^{\ast AIPW})$ \\ 
  \hline
(-1,-1) & 2.234 & 2.194 & 0.458 & 2.198 & 0.464 & 69.284 & 6.232 & 0.014 & 0.016 \\ 
  (-1,1) & 2.758 & 2.763 & 0.364 & 2.764 & 0.352 & 80.708 & 6.698 & 0.176 & 0.158 \\ 
  (1,-1) & 1.76 & 1.806 & 0.586 & 1.773 & 0.515 & 70.104 & 6.324 & 0.012 & 0 \\ 
  (1,1) & 3.251 & 3.249 & 0.391 & 3.25 & 0.381 & 80.302 & 6.882 & 0.798 & 0.826 \\
   \hline
\multicolumn{2}{c}{DTR} &
  \multicolumn{8}{c}{SMART-EXAM $(C=0.6)$, Positive} \\
  \cmidrule(lr){1-2}\cmidrule(lr){3-10}
DTR & $\mu_{\bm{d}_j}$ & 
$E[\hat{\mu}^{IPW}_{\bm{d}_j}]$ & $\se (\hat{\mu}^{IPW}_{\bm{d}_j})$ & $E[\hat{\mu}^{AIPW}_{\bm{d}_j}]$ & $\se (\hat{\mu}^{AIPW}_{\bm{d}_j})$ & $\bar{N}_{\bm{d}_j}$ & $\se (N_{\bm{d}_j})$ & $\Pr(\bm{d}_j=\hat{\bm{d}}^{\ast IPW})$ & $\Pr(\bm{d}_j=\hat{\bm{d}}^{\ast AIPW})$ \\ 
  \hline
(-1,-1) & 2.234 & 2.194 & 0.452 & 2.194 & 0.458 & 69.394 & 6.26 & 0.014 & 0.018 \\ 
  (-1,1) & 2.758 & 2.764 & 0.372 & 2.762 & 0.359 & 80.598 & 6.721 & 0.178 & 0.156 \\ 
  (1,-1) & 1.76 & 1.785 & 0.572 & 1.758 & 0.522 & 69.598 & 6.353 & 0.004 & 0 \\ 
  (1,1) & 3.251 & 3.251 & 0.377 & 3.247 & 0.367 & 80.808 & 6.833 & 0.804 & 0.826 \\
   \hline
   
 \multicolumn{2}{c}{DTR} &
  \multicolumn{8}{c}{SMART-EXAM $(C=0.7)$, Negative} \\
  \cmidrule(lr){1-2}\cmidrule(lr){3-10}
DTR & $\mu_{\bm{d}_j}$ & 
$E[\hat{\mu}^{IPW}_{\bm{d}_j}]$ & $\se (\hat{\mu}^{IPW}_{\bm{d}_j})$ & $E[\hat{\mu}^{AIPW}_{\bm{d}_j}]$ & $\se (\hat{\mu}^{AIPW}_{\bm{d}_j})$ & $\bar{N}_{\bm{d}_j}$ & $\se (N_{\bm{d}_j})$ & $\Pr(\bm{d}_j=\hat{\bm{d}}^{\ast IPW})$ & $\Pr(\bm{d}_j=\hat{\bm{d}}^{\ast AIPW})$ \\   
  \hline
(-1,-1) & 2.234 & 2.208 & 0.483 & 2.208 & 0.494 & 64.946 & 6.067 & 0.01 & 0.018 \\ 
  (-1,1) & 2.758 & 2.753 & 0.338 & 2.75 & 0.335 & 85.046 & 6.935 & 0.152 & 0.152 \\ 
  (1,-1) & 1.76 & 1.8 & 0.603 & 1.77 & 0.55 & 65.758 & 6.533 & 0.008 & 0 \\ 
  (1,1) & 3.251 & 3.265 & 0.356 & 3.264 & 0.347 & 84.648 & 6.866 & 0.83 & 0.83 \\ 
   \hline
   \multicolumn{2}{c}{DTR} &
  \multicolumn{8}{c}{SMART-EXAM $(C=0.7)$, Positive} \\
  \cmidrule(lr){1-2}\cmidrule(lr){3-10}
DTR & $\mu_{\bm{d}_j}$ & 
$E[\hat{\mu}^{IPW}_{\bm{d}_j}]$ & $\se (\hat{\mu}^{IPW}_{\bm{d}_j})$ & $E[\hat{\mu}^{AIPW}_{\bm{d}_j}]$ & $\se (\hat{\mu}^{AIPW}_{\bm{d}_j})$ & $\bar{N}_{\bm{d}_j}$ & $\se (N_{\bm{d}_j})$ & $\Pr(\bm{d}_j=\hat{\bm{d}}^{\ast IPW})$ & $\Pr(\bm{d}_j=\hat{\bm{d}}^{\ast AIPW})$ \\   
  \hline
(-1,-1) & 2.234 & 2.204 & 0.466 & 2.202 & 0.475 & 65.098 & 6.052 & 0.018 & 0.026 \\ 
  (-1,1) & 2.758 & 2.756 & 0.345 & 2.754 & 0.34 & 84.894 & 6.949 & 0.142 & 0.138 \\ 
  (1,-1) & 1.76 & 1.789 & 0.607 & 1.753 & 0.562 & 64.87 & 6.553 & 0.004 & 0.002 \\ 
  (1,1) & 3.251 & 3.266 & 0.354 & 3.267 & 0.347 & 85.536 & 6.945 & 0.836 & 0.834 \\ 
   \hline
   \multicolumn{2}{c}{DTR} &
  \multicolumn{8}{c}{SMART-AR-EXAM $(C=0.5)$, Negative}  \\
  \cmidrule(lr){1-2}\cmidrule(lr){3-10}
DTR & $\mu_{\bm{d}_j}$ & 
$E[\hat{\mu}^{IPW}_{\bm{d}_j}]$ & $\se (\hat{\mu}^{IPW}_{\bm{d}_j})$ & $E[\hat{\mu}^{AIPW}_{\bm{d}_j}]$ & $\se (\hat{\mu}^{AIPW}_{\bm{d}_j})$ & $\bar{N}_{\bm{d}_j}$ & $\se (N_{\bm{d}_j})$ & $\Pr(\bm{d}_j=\hat{\bm{d}}^{\ast IPW})$ & $\Pr(\bm{d}_j=\hat{\bm{d}}^{\ast AIPW})$ \\  
  \hline
(-1,-1) & 2.234 & 2.26 & 0.384 & 2.263 & 0.387 & 74.632 & 6.69 & 0.006 & 0.006 \\ 
  (-1,1) & 2.758 & 2.721 & 0.392 & 2.726 & 0.393 & 75.36 & 6.423 & 0.156 & 0.164 \\ 
  (1,-1) & 1.76 & 1.799 & 0.489 & 1.775 & 0.454 & 74.932 & 6.548 & 0.002 & 0 \\ 
  (1,1) & 3.251 & 3.26 & 0.409 & 3.25 & 0.408 & 75.474 & 6.648 & 0.836 & 0.83 \\ 
   \hline
     \multicolumn{2}{c}{DTR} &
  \multicolumn{8}{c}{SMART-AR-EXAM $(C=0.5)$, Positive}  \\
  \cmidrule(lr){1-2}\cmidrule(lr){3-10}
DTR & $\mu_{\bm{d}_j}$ & 
$E[\hat{\mu}^{IPW}_{\bm{d}_j}]$ & $\se (\hat{\mu}^{IPW}_{\bm{d}_j})$ & $E[\hat{\mu}^{AIPW}_{\bm{d}_j}]$ & $\se (\hat{\mu}^{AIPW}_{\bm{d}_j})$ & $\bar{N}_{\bm{d}_j}$ & $\se (N_{\bm{d}_j})$ & $\Pr(\bm{d}_j=\hat{\bm{d}}^{\ast IPW})$ & $\Pr(\bm{d}_j=\hat{\bm{d}}^{\ast AIPW})$ \\  
  \hline
(-1,-1) & 2.234 & 2.257 & 0.388 & 2.256 & 0.388 & 74.664 & 6.638 & 0.01 & 0.006 \\ 
  (-1,1) & 2.758 & 2.729 & 0.395 & 2.737 & 0.39 & 75.328 & 6.472 & 0.176 & 0.188 \\ 
  (1,-1) & 1.76 & 1.816 & 0.483 & 1.78 & 0.45 & 74.568 & 6.468 & 0 & 0 \\ 
  (1,1) & 3.251 & 3.248 & 0.406 & 3.245 & 0.406 & 75.838 & 6.72 & 0.814 & 0.806 \\
   \hline
\end{tabular}
}
\end{table}

\begin{table}
 \centering
 \caption{ The operating characteristics of each SMART design when the total sample size is $N=200$, the sample size for the pilot SMART is $N^p=200$, $\epsilon=0.3$, and $\eta=-1$. $\mu_{\bm{d}_j}$ denotes the true value of DTR $\bm{d}_j$;  $E[\hat{\mu}^{IPW}_{\bm{d}_j}]$ and $E[\hat{\mu}^{AIPW}_{\bm{d}_j}]$ denote the Monte Carlo mean of the estimated DTR mean by IPW and AIPW, respectively;  $\se (\hat{\mu}^{IPW}_{\bm{d}_j})$ and $\se (\hat{\mu}^{AIPW}_{\bm{d}_j})$ denote the empirical standard error of the estimated DTR mean by IPW and AIPW, respectively; $\bar{N}_{\bm{d}_j}$ is the Monte Carlo mean of the number of participants in DTR $\bm{d}_j$; $\se (N_{\bm{d}_j})$ denotes the empirical standard error of the number of participants in DTR $\bm{d}_j$; $\Pr(\bm{d}_j=\hat{\bm{d}}^{\ast IPW})$ and $\Pr(\bm{d}_j=\hat{\bm{d}}^{\ast AIPW})$ correspond to the probability of being selected as the optimal DTR for each DTR $\bm{d}_j$ by IPW and AIPW, respectively.}
 \label{table_3}
 \resizebox{0.7\columnwidth}{!}{%
\begin{tabular}{lccccccccc}
  \hline
   \multicolumn{2}{c}{DTR} & \multicolumn{8}{c}{SMART, Negative}  \\
  \cmidrule(lr){1-2}\cmidrule(lr){3-10}
DTR & $\mu_{\bm{d}_j}$ & 
$E[\hat{\mu}^{IPW}_{\bm{d}_j}]$ & $\se (\hat{\mu}^{IPW}_{\bm{d}_j})$ & $E[\hat{\mu}^{AIPW}_{\bm{d}_j}]$ & $\se (\hat{\mu}^{AIPW}_{\bm{d}_j})$ & $\bar{N}_{\bm{d}_j}$ & $\se (N_{\bm{d}_j})$ & $\Pr(\bm{d}_j=\hat{\bm{d}}^{\ast IPW})$ & $\Pr(\bm{d}_j=\hat{\bm{d}}^{\ast AIPW})$ \\ 
  \hline
(-1,-1) & 2.234 & 2.22 & 0.36 & 2.222 & 0.362 & 74.936 & 6.654 & 0.008 & 0.008 \\ 
  (-1,1) & 2.758 & 2.756 & 0.375 & 2.759 & 0.375 & 75.056 & 6.848 & 0.166 & 0.166 \\ 
  (1,-1) & 1.76 & 1.777 & 0.461 & 1.766 & 0.429 & 75.244 & 6.739 & 0.002 & 0 \\ 
  (1,1) & 3.251 & 3.28 & 0.382 & 3.274 & 0.382 & 75.162 & 6.816 & 0.824 & 0.826 \\ 
   \hline
   \multicolumn{2}{c}{DTR} & \multicolumn{8}{c}{SMART, Positive}  \\
  \cmidrule(lr){1-2}\cmidrule(lr){3-10}
DTR & $\mu_{\bm{d}_j}$ & 
$E[\hat{\mu}^{IPW}_{\bm{d}_j}]$ & $\se (\hat{\mu}^{IPW}_{\bm{d}_j})$ & $E[\hat{\mu}^{AIPW}_{\bm{d}_j}]$ & $\se (\hat{\mu}^{AIPW}_{\bm{d}_j})$ & $\bar{N}_{\bm{d}_j}$ & $\se (N_{\bm{d}_j})$ & $\Pr(\bm{d}_j=\hat{\bm{d}}^{\ast IPW})$ & $\Pr(\bm{d}_j=\hat{\bm{d}}^{\ast AIPW})$ \\ 
  \hline
 (-1,-1) & 2.234 & 2.22 & 0.36 & 2.222 & 0.362 & 74.936 & 6.654 & 0.008 & 0.008 \\ 
  (-1,1) & 2.758 & 2.756 & 0.375 & 2.759 & 0.375 & 75.056 & 6.848 & 0.166 & 0.166 \\ 
  (1,-1) & 1.76 & 1.777 & 0.461 & 1.766 & 0.429 & 75.244 & 6.739 & 0.002 & 0 \\ 
  (1,1) & 3.251 & 3.28 & 0.382 & 3.274 & 0.382 & 75.162 & 6.816 & 0.824 & 0.826 \\ 
  \hline
  \multicolumn{2}{c}{DTR} &
  \multicolumn{8}{c}{SMART-EXAM $(C=0.5)$, Negative} \\
  \cmidrule(lr){1-2}\cmidrule(lr){3-10}
DTR & $\mu_{\bm{d}_j}$ & 
$E[\hat{\mu}^{IPW}_{\bm{d}_j}]$ & $\se (\hat{\mu}^{IPW}_{\bm{d}_j})$ & $E[\hat{\mu}^{AIPW}_{\bm{d}_j}]$ & $\se (\hat{\mu}^{AIPW}_{\bm{d}_j})$ & $\bar{N}_{\bm{d}_j}$ & $\se (N_{\bm{d}_j})$ & $\Pr(\bm{d}_j=\hat{\bm{d}}^{\ast IPW})$ & $\Pr(\bm{d}_j=\hat{\bm{d}}^{\ast AIPW})$ \\ 
  \hline
(-1,-1) & 2.234 & 2.231 & 0.389 & 2.233 & 0.387 & 74.248 & 6.441 & 0.014 & 0.012 \\ 
  (-1,1) & 2.758 & 2.751 & 0.378 & 2.753 & 0.37 & 75.744 & 6.734 & 0.172 & 0.172 \\ 
  (1,-1) & 1.76 & 1.792 & 0.504 & 1.77 & 0.464 & 74.602 & 6.525 & 0.002 & 0.004 \\ 
  (1,1) & 3.251 & 3.255 & 0.418 & 3.253 & 0.409 & 75.804 & 6.848 & 0.812 & 0.812 \\
   \hline
\multicolumn{2}{c}{DTR} &
  \multicolumn{8}{c}{SMART-EXAM $(C=0.5)$, Positive} \\
  \cmidrule(lr){1-2}\cmidrule(lr){3-10}
DTR & $\mu_{\bm{d}_j}$ & 
$E[\hat{\mu}^{IPW}_{\bm{d}_j}]$ & $\se (\hat{\mu}^{IPW}_{\bm{d}_j})$ & $E[\hat{\mu}^{AIPW}_{\bm{d}_j}]$ & $\se (\hat{\mu}^{AIPW}_{\bm{d}_j})$ & $\bar{N}_{\bm{d}_j}$ & $\se (N_{\bm{d}_j})$ & $\Pr(\bm{d}_j=\hat{\bm{d}}^{\ast IPW})$ & $\Pr(\bm{d}_j=\hat{\bm{d}}^{\ast AIPW})$ \\ 
  \hline
(-1,-1) & 2.234 & 2.218 & 0.382 & 2.219 & 0.38 & 74.368 & 6.58 & 0.006 & 0.008 \\ 
  (-1,1) & 2.758 & 2.767 & 0.384 & 2.768 & 0.375 & 75.624 & 6.683 & 0.162 & 0.186 \\ 
  (1,-1) & 1.76 & 1.768 & 0.514 & 1.75 & 0.475 & 74.58 & 6.547 & 0.002 & 0 \\ 
  (1,1) & 3.251 & 3.275 & 0.397 & 3.27 & 0.393 & 75.826 & 6.956 & 0.83 & 0.806 \\ 
   \hline
   
\multicolumn{2}{c}{DTR} &
  \multicolumn{8}{c}{SMART-EXAM $(C=0.6)$, Negative} \\
  \cmidrule(lr){1-2}\cmidrule(lr){3-10}
DTR & $\mu_{\bm{d}_j}$ & 
$E[\hat{\mu}^{IPW}_{\bm{d}_j}]$ & $\se (\hat{\mu}^{IPW}_{\bm{d}_j})$ & $E[\hat{\mu}^{AIPW}_{\bm{d}_j}]$ & $\se (\hat{\mu}^{AIPW}_{\bm{d}_j})$ & $\bar{N}_{\bm{d}_j}$ & $\se (N_{\bm{d}_j})$ & $\Pr(\bm{d}_j=\hat{\bm{d}}^{\ast IPW})$ & $\Pr(\bm{d}_j=\hat{\bm{d}}^{\ast AIPW})$ \\ 
  \hline
(-1,-1) & 2.234 & 2.218 & 0.42 & 2.217 & 0.425 & 69.176 & 6.199 & 0.012 & 0.016 \\ 
  (-1,1) & 2.758 & 2.756 & 0.345 & 2.753 & 0.338 & 80.816 & 6.928 & 0.156 & 0.154 \\ 
  (1,-1) & 1.76 & 1.799 & 0.525 & 1.77 & 0.479 & 70.032 & 6.609 & 0.002 & 0 \\ 
  (1,1) & 3.251 & 3.26 & 0.377 & 3.263 & 0.367 & 80.374 & 7.032 & 0.83 & 0.83 \\ 
   \hline
\multicolumn{2}{c}{DTR} &
  \multicolumn{8}{c}{SMART-EXAM $(C=0.6)$, Positive} \\
  \cmidrule(lr){1-2}\cmidrule(lr){3-10}
DTR & $\mu_{\bm{d}_j}$ & 
$E[\hat{\mu}^{IPW}_{\bm{d}_j}]$ & $\se (\hat{\mu}^{IPW}_{\bm{d}_j})$ & $E[\hat{\mu}^{AIPW}_{\bm{d}_j}]$ & $\se (\hat{\mu}^{AIPW}_{\bm{d}_j})$ & $\bar{N}_{\bm{d}_j}$ & $\se (N_{\bm{d}_j})$ & $\Pr(\bm{d}_j=\hat{\bm{d}}^{\ast IPW})$ & $\Pr(\bm{d}_j=\hat{\bm{d}}^{\ast AIPW})$ \\ 
  \hline
(-1,-1) & 2.234 & 2.216 & 0.41 & 2.217 & 0.416 & 69.232 & 6.286 & 0.014 & 0.016 \\ 
  (-1,1) & 2.758 & 2.758 & 0.354 & 2.755 & 0.349 & 80.76 & 6.938 & 0.148 & 0.14 \\ 
  (1,-1) & 1.76 & 1.789 & 0.53 & 1.765 & 0.5 & 69.452 & 6.558 & 0.002 & 0 \\ 
  (1,1) & 3.251 & 3.265 & 0.36 & 3.266 & 0.352 & 80.954 & 7.031 & 0.836 & 0.844 \\ 
   \hline
   
 \multicolumn{2}{c}{DTR} &
  \multicolumn{8}{c}{SMART-EXAM $(C=0.7)$, Negative} \\
  \cmidrule(lr){1-2}\cmidrule(lr){3-10}
DTR & $\mu_{\bm{d}_j}$ & 
$E[\hat{\mu}^{IPW}_{\bm{d}_j}]$ & $\se (\hat{\mu}^{IPW}_{\bm{d}_j})$ & $E[\hat{\mu}^{AIPW}_{\bm{d}_j}]$ & $\se (\hat{\mu}^{AIPW}_{\bm{d}_j})$ & $\bar{N}_{\bm{d}_j}$ & $\se (N_{\bm{d}_j})$ & $\Pr(\bm{d}_j=\hat{\bm{d}}^{\ast IPW})$ & $\Pr(\bm{d}_j=\hat{\bm{d}}^{\ast AIPW})$ \\   
  \hline
(-1,-1) & 2.234 & 2.239 & 0.449 & 2.235 & 0.472 & 64.852 & 6.218 & 0.018 & 0.018 \\ 
  (-1,1) & 2.758 & 2.742 & 0.332 & 2.743 & 0.334 & 85.14 & 6.932 & 0.126 & 0.126 \\ 
  (1,-1) & 1.76 & 1.763 & 0.565 & 1.746 & 0.519 & 65.276 & 6.542 & 0.002 & 0 \\ 
  (1,1) & 3.251 & 3.284 & 0.337 & 3.282 & 0.333 & 85.13 & 7.033 & 0.854 & 0.856 \\
   \hline
   \multicolumn{2}{c}{DTR} &
  \multicolumn{8}{c}{SMART-EXAM $(C=0.7)$, Positive} \\
  \cmidrule(lr){1-2}\cmidrule(lr){3-10}
DTR & $\mu_{\bm{d}_j}$ & 
$E[\hat{\mu}^{IPW}_{\bm{d}_j}]$ & $\se (\hat{\mu}^{IPW}_{\bm{d}_j})$ & $E[\hat{\mu}^{AIPW}_{\bm{d}_j}]$ & $\se (\hat{\mu}^{AIPW}_{\bm{d}_j})$ & $\bar{N}_{\bm{d}_j}$ & $\se (N_{\bm{d}_j})$ & $\Pr(\bm{d}_j=\hat{\bm{d}}^{\ast IPW})$ & $\Pr(\bm{d}_j=\hat{\bm{d}}^{\ast AIPW})$ \\   
  \hline
(-1,-1) & 2.234 & 2.239 & 0.449 & 2.235 & 0.472 & 64.852 & 6.218 & 0.018 & 0.018 \\ 
  (-1,1) & 2.758 & 2.742 & 0.332 & 2.743 & 0.334 & 85.14 & 6.932 & 0.126 & 0.126 \\ 
  (1,-1) & 1.76 & 1.763 & 0.565 & 1.746 & 0.519 & 65.276 & 6.542 & 0.002 & 0 \\ 
  (1,1) & 3.251 & 3.284 & 0.337 & 3.282 & 0.333 & 85.13 & 7.033 & 0.854 & 0.856 \\ 
   \hline
   \multicolumn{2}{c}{DTR} &
  \multicolumn{8}{c}{SMART-AR-EXAM $(C=0.5)$, Negative}  \\
  \cmidrule(lr){1-2}\cmidrule(lr){3-10}
DTR & $\mu_{\bm{d}_j}$ & 
$E[\hat{\mu}^{IPW}_{\bm{d}_j}]$ & $\se (\hat{\mu}^{IPW}_{\bm{d}_j})$ & $E[\hat{\mu}^{AIPW}_{\bm{d}_j}]$ & $\se (\hat{\mu}^{AIPW}_{\bm{d}_j})$ & $\bar{N}_{\bm{d}_j}$ & $\se (N_{\bm{d}_j})$ & $\Pr(\bm{d}_j=\hat{\bm{d}}^{\ast IPW})$ & $\Pr(\bm{d}_j=\hat{\bm{d}}^{\ast AIPW})$ \\  
  \hline
(-1,-1) & 2.234 & 2.253 & 0.377 & 2.255 & 0.378 & 74.684 & 6.759 & 0.01 & 0.01 \\ 
  (-1,1) & 2.758 & 2.729 & 0.383 & 2.734 & 0.38 & 75.308 & 6.502 & 0.152 & 0.164 \\ 
  (1,-1) & 1.76 & 1.787 & 0.469 & 1.773 & 0.441 & 74.996 & 6.624 & 0.004 & 0 \\ 
  (1,1) & 3.251 & 3.259 & 0.39 & 3.25 & 0.392 & 75.41 & 6.707 & 0.834 & 0.826 \\ 
   \hline
     \multicolumn{2}{c}{DTR} &
  \multicolumn{8}{c}{SMART-AR-EXAM $(C=0.5)$, Positive}  \\
  \cmidrule(lr){1-2}\cmidrule(lr){3-10}
DTR & $\mu_{\bm{d}_j}$ & 
$E[\hat{\mu}^{IPW}_{\bm{d}_j}]$ & $\se (\hat{\mu}^{IPW}_{\bm{d}_j})$ & $E[\hat{\mu}^{AIPW}_{\bm{d}_j}]$ & $\se (\hat{\mu}^{AIPW}_{\bm{d}_j})$ & $\bar{N}_{\bm{d}_j}$ & $\se (N_{\bm{d}_j})$ & $\Pr(\bm{d}_j=\hat{\bm{d}}^{\ast IPW})$ & $\Pr(\bm{d}_j=\hat{\bm{d}}^{\ast AIPW})$ \\  
  \hline
(-1,-1) & 2.234 & 2.249 & 0.371 & 2.252 & 0.371 & 74.71 & 6.734 & 0.012 & 0.016 \\ 
  (-1,1) & 2.758 & 2.733 & 0.379 & 2.743 & 0.378 & 75.282 & 6.55 & 0.158 & 0.16 \\ 
  (1,-1) & 1.76 & 1.801 & 0.472 & 1.772 & 0.445 & 74.606 & 6.47 & 0 & 0 \\ 
  (1,1) & 3.251 & 3.25 & 0.382 & 3.247 & 0.384 & 75.8 & 6.837 & 0.83 & 0.824 \\ 
   \hline
\end{tabular}
}
\end{table}

\begin{table}
    \centering
    \caption{Parameter specification for generating data from each SMART design in the empirical application. }
    \label{table_4}
    \resizebox{.7\textwidth}{!}{
    \begin{tabularx}{\textwidth}{sq}
    \hline
     Variable & Specification \\
    \hline
 \multirow{4}{*}{Baseline variables}  & $O_{11} \sim \ber(0.35)$ \\
       & $O_{12} \sim N(-0.12, 1)$ \\
       & $O_{13} \sim \ber(0.31)$ \\
       & $O_{14} \sim \ber(0.81)$ \\
       Response status  & The response rates for treatment $A_1=1$ and $A_1=-1$ are specified as $\pi_1=0.31$ and $\pi_{-1}=0.37$, respectively. \\
       \multirow{2}{*}{Tailoring variable} & $O_{22} \sim \ber(0.42)$ for those assigned to treatment $A_2=1$ \\
       & $O_{22} \sim \ber(0.53)$ for those assigned to treatment $A_2=-1$\\
       \multirow{3}{*}{Preference indicator $\Lambda_i$}  & S1: $\pr(\Lambda_{i}=1 | O_{22,i}=0) = 0.8$ and $\pr(\Lambda_{i}=1 | O_{22,i}=1) = 0.4$ \\
       & S2: $\pr(\Lambda_{i}=1 | O_{22,i}=0) = 0.4$ and $\pr(\Lambda_{i}=1 | O_{22,i}=1) = 0.3$ \\
       & S3: $\pr(\Lambda_{i}=1 | O_{22,i}=0) = 0.8$ and $\pr(\Lambda_{i}=1 | O_{22,i}=1) = 0.6$ \\
       \multirow{2}{*}{The outcome $Y$} & For non-responders:
$Y_i =2.69-0.25 O_{11,i}-0.30 O_{12,i} + 0.04 O_{13,i} + 0.49 O_{14,i} + 0.08 A_{1i} - 0.09 O_{22,i} + 0.86 A_{2i} + 0.19 A_{1i} A_{2i} - 1.18 O_{22,i} A_{2i} + \tau, \quad \tau \sim N(0,1)$ \\
 & For responders:
    $Y_i=3.00-0.62 O_{11,i}-0.41 O_{12,i} - 0.10 O_{13,i} + 0.38 O_{14,i} + 0.10 A_{1i}  + \tau, \quad \tau \sim N(0,1).$ \\
      Treatment capacity & Equal treatment capacity $C=0.5$ \\
  \hline
    \end{tabularx}
    }
\end{table}